\documentclass [aps,preprint,prd,showpacs,nofootinbib,oneside]{revtex4}
\usepackage{verbatim}%加注释行用的%
\usepackage{slashed}%输入slash时加载的。 例子：（/slashed{p}）%
\usepackage{epsfig}%加入eps的图形文件%
\usepackage{feynmf}
\usepackage{graphicx}%加入一般图形文件%
\usepackage{amsmath}%数学公式输入宏包%
\usepackage{mathrsfs}%花写字体的输入%
\usepackage{bm}

\begin{document}

%\preprint{}

\title{NRQCD Predictions of D-Wave Quarkonia $^3D_{J}(J=1,2,3)$
Decay into Light Hadrons at Order $\alpha_{s}^{3}$}


\author{Zhi-Guo He\footnote{Present address: Departament d'Estructura i
Constituents de la Mat\`eria and Institut de Ci\`encies del
Cosmos,Universitat de Barcelona, Diagonal, 647, E-08028 Barcelona,
Catalonia, Spain. }}
%\email{hzgzlh@gmail.com}
\author{Ying \surname{Fan}}
%\email{ying.physics.fan@gamil.com}
\author{Kuang-Ta Chao}
%\email{ktchao@th.phy.pku.edu.cn}
\affiliation{Department of Physics
and State Key Laboratory of Nuclear Physics and Technology, Peking
University, Beijing 100871, China}

%\affiliation{China Center of Advanced Science and Technology (World
%Laboratory), Beijing 100080, China}

%\thanks{}
%\affiliation{}


%\date{\today}

\begin{abstract}
In this paper, in the framework of nonrelativistic QCD  we
study the light hadron (LH) decays of the spin-triplet ($S$=1) $D$-wave
heavy quarkonia. The short-distance coefficients of all Fock states
in the $^3D_J(J=1,2,3)$ quarkonia including the $D$-wave color singlet,
$P$-wave color octet, and $S$-wave color singlet and color octet are
calculated perturbatively at $\alpha_{s}^3$ order.  The operator
evolution equations of the four-fermion operators are also derived
and are used to estimate the numerical values of the long-distance
matrix elements. We find that for the $c\bar{c}$ system, the LH
decay widths of $\psi(1^3D_J)$ predicted by nonrelativistic QCD
is about $2\sim3$ times larger than the phenomenological potential
model results, while for the $b\bar{b}$ system the two theoretical
estimations of $\Gamma(\Upsilon(1^3D_J)\to \mathrm{LH})$  are in coincidence
with each other. Our predictions for $\psi(1^3D_J)$ LH decay widths
are $\Gamma(\psi(1^3D_J)\to \mathrm{LH})=(435,50,172)$keV for $J$=1,2,3; and
for $\Upsilon(1^3D_J)$, $\Gamma(\Upsilon(1^3D_J)\to
\mathrm{LH})=(6.91,0.75,2.75)$keV for $J$=1,2,3.
\end{abstract}

\pacs{12.38.Bx, 12.39.Jh, 13.20.Gd}

\maketitle

%\keywords{}



\section{Introduction}
   The production, decay, and mass spectrum of heavy quarkonium have
been interesting topics since the first charmonium state $J/\psi$
was discovered in 1974. Because of their large mass scales and
nonrelativistic nature, heavy quarkonia are good probes to study
and understand quantum chromodynamics (QCD) from both perturbative
and nonperturbative aspects. In fact, one of the earliest
applications of QCD is to calculate the inclusive decay rates of
heavy quarkonia. In early times, it was assumed that such a decay
process can proceed through two steps. First, the heavy quarkonium
transforms into a free $Q\bar{Q}$ pair, which is a long-distance
nonperturbative effect. Then, the heavy-quark pair annihilates into
light hadrons (LH) through gluons, which can be calculated
perturbatively. In the nonrelativistic limit, the long-distance
part is related to the $Q\bar{Q}$ Schr\"odinger wave functions or
their derivatives at the origin. In this picture, the free
$Q\bar{Q}$ are in color singlet and have the same quantum numbers
$J^{PC}$ as the bound state heavy quarkonium. This is referred to as
the "color-singlet model". Explicit calculations at
next-to-leading order (NLO) in $\alpha_{s}$ for $S$-wave quarkonium
decays support the color-singlet model factorization formula. But it
breaks down in the calculations of $P$-wave \cite{Barbieri:1976fp,BBL6}
and $D$-wave\cite{L.Berg,G.Belanger} heavy quarkonium LH decays at
$\alpha_{s}^{3}$ order, where infrared divergences appear.
Phenomenologically, these infrared divergences are regularized by
the binding energy of $Q\bar{Q}$  bound states.

  In Ref.\cite{BBL1}, Bodwin, Braaten, and Lepage first introduced
the color-octet matrix elements to absorb the infrared logarithms,
then they developed nonrelativistic QCD (NRQCD) effective field
theory\cite{BBL2}, based on which the inclusive decay rate of heavy
quarkonium can be given by a rigorous factorization formula, and
calculated in a systematic way by double expansion of $\alpha_{s}$,
the coupling constant of QCD, and $v$, the typical velocity of heavy
quarks in the heavy quarkonium. In their formula, heavy quarkonium is
treated as a superposition state of $|Q\bar{Q}\rangle, |Q\bar{Q}g\rangle,
|Q\bar{Q}gg\rangle$, and other higher order Fock states, rather than
 the $|Q\bar{Q}\rangle$ color-singlet state only. The contribution of
each Fock state is organized in powers of $v^{2}$, and can be
written as a product of the long-distance matrix element and the
corresponding short-distance coefficient. Huang and
Chao\cite{Huang1} first got the infrared finite LH decay width of
the spin singlet $P$-wave state, $h_{c}$, with QCD radiative corrections
in the framework of NRQCD. The decay widths of
$\chi_{c_{J}}\rightarrow \mathrm{LH}$ were calculated to $\alpha_{s}^{3}$
order in Refs.\cite{Huang3,Petrelli}. Complete and detailed results
of color-singlet and octet-short distance coefficients of $S$-wave and
$P$-wave spin-triplet states were given in Ref.\cite{maltoni}. In
Refs.\cite{Huang2,maltoni}, the authors also explained why the
infrared divergences disappear in the NRQCD factorization approach.

NRQCD is now a widely accepted effective field theory for heavy
quarkonium. In the framework of NRQCD,  lots of theoretical work
has been done to study $S$- and $P$-wave quarkonium decays, and some
significant successes have been achieved (for a review see
Ref.\cite{Brambilla:2004wf}). Recently, the order $v^{7}$ results of
$S$- and $P$-wave heavy quarkonium inclusive hadronic decays were
obtained by Brambilla et al.\cite{Brambilla:2008zg} However,
compared with $S$- and $P$-wave heavy quarkonium, little work has been
done on $D$-wave states in NRQCD. In this paper, we will calculate the
LH decay widths of spin-triplet $D$-wave states $^3D_{J}$ ($J$=1,2,3).
Here the subprocesses $^3S_{1}^{[1,8]}\rightarrow \mathrm{LH}$ and
$^3P_{J}^{[8]}\rightarrow \mathrm{LH}$ at leading-order (LO) in $v^{2}$ are
all included, in which the short-distance coefficients are
calculated in a different way and in agreement with the results in
Ref.\cite{maltoni}. As the main part of these $D$-wave quarkonium
decays, the infrared safe short-distance coefficients of
$^3D_{J}^{[1]}\rightarrow \mathrm{LH}$ are first obtained in this paper. We
use the covariant projection method, which was first introduced
in\cite{Ku} and generalized in Refs.\cite{maltoni,v4}, to do the
perturbative calculations. At LO in $v^{2}$, the long-distance
matrix elements of color-singlet $D$-wave four-fermion operators are
related to the wave function's second derivative at the origin. And
the matrix elements of the $S$-wave and $P$-wave octet and of $S$-wave singlet
four-fermion operators could be studied in lattice simulations, or
determined by fitting experimental data, or roughly estimated
through velocity scaling rules. To give numerical predictions, here
we use operator evolution equations to estimate the values of the
matrix elements. The rest of our paper is organized as follows: In
Sec.II, we briefly introduce the NRQCD effective field theory
and give the general formulas used in $D$ dimension. In Sec.III,
all the subprocesses will be calculated to $\alpha_{s}^{3}$ order.
And after matching the full QCD results with the NLO NRQCD ones, the
infrared safe short-distance coefficients as well as the operator
evolution equations are obtained in Sec.IV. In Sec.V, we
will discuss the numerical results and their phenomenological
applications to $c\bar{c}$ and $b\bar{b}$ systems.

\section{General Formulas}

  The Lagrangian for NRQCD is\cite{BBL2}:
\begin{equation}
\mathcal{L}_{NRQCD}=\mathcal{L}_{light}+\mathcal{L}_{heavy}+\delta\mathcal{L},
\end{equation}
where $\mathcal{L}_{light}$ includes gauge field and light quark
parts, and $\mathcal{L}_{heavy}$ is the nonrelativistic Lagrangian
for the heavy quarks and antiquarks:
\begin{equation}\label{Lag}
\mathcal{L}_{heavy}=\psi^{\dagger}(iD_{t}+\frac{\mathbf{D}^{2}}{2m_{Q}})\psi+
\chi^{\dagger}(iD_{t}-\frac{\mathbf{D}^{2}}{2m_{Q}})\chi,
\end{equation}
where $\psi$ and $\chi^{\dagger}$ are the Pauli spinor fields that
annihilate heavy quark and antiquark, respectively, and $D_{t}$ and
$\mathbf{D}$ are the time and space components of the
gauge-covariant derivative $D^{\mu}$. $\delta\mathcal{L}$ describes
the relativistic effects. The leading-order $v^{2}$ corrections are
the bilinear terms:
\begin{eqnarray}
\delta\mathcal{L}_{bilinear}&=&\frac{c_{1}}{8m_{Q}^3}[\psi^{\dagger}(\mathbf{D}^2)^{2}\psi-
\chi^{\dagger}(\mathbf{D}^2)^{2}\chi]\nonumber\\
&+&\frac{c_{2}}{8m_{Q}^2} [\psi^{\dagger}(\mathbf{D}\cdot
g\mathbf{E}-\mathbf{E}\cdot
g\mathbf{D})\psi+\chi^{\dagger}(\mathbf{D}\cdot
g\mathbf{E}-\mathbf{E}\cdot g\mathbf{D})\chi]\nonumber \\
&+&\frac{c_{3}}{8m_{Q}^2}[\psi^{\dagger}(i\mathbf{D}\times
g\mathbf{E}-g\mathbf{E}\times
i\mathbf{D})\cdot\boldsymbol{\sigma}\psi+\chi^{\dagger}(i\mathbf{D}\times
g\mathbf{E}-g\mathbf{E}\times
i\mathbf{D})\cdot\boldsymbol{\sigma}\chi]\nonumber\\
&+&\frac{c_{4}}{2m_{Q}}[\psi^{\dagger}(g\mathbf{B}\cdot\boldsymbol{\sigma})\psi-
\chi^{\dagger}(g\mathbf{B}\cdot\boldsymbol{\sigma})\chi],
\end{eqnarray}
where $c_{i}$ could be obtained by using the Foldy-Wouthuysen-Tani
transformation\cite{FWT1}, diagonalizing the Dirac theory so as to
decouple the heavy-quark and antiquark degrees of freedom. To
reproduce the full QCD Lagrangian and to describe the annihilation
of the heavy-quark pair, four-fermion local operator terms are also
needed:
\begin{equation}
\delta\mathcal{L}_{4-fermion}=\sum_{n}\frac{f_{n}(\mu_\Lambda)}{m_{Q}^{d_{n}-4}}\mathcal{O}_{n}(\mu_\Lambda),
\end{equation}
where $\mathcal{O}_{n}(\mu_\Lambda)$ is the local four-fermion
operator with the general form
$(\psi^{\dagger}\mathcal{K}_{i}\chi)(\chi^{\dagger}\mathcal{K}^{\prime}_{i}\psi)$,
and $f_{n}(\mu_\Lambda)$ is the Wilson coefficient. In effective
theory both operators and coefficients are dependent on the
factorization scale $\mu_\Lambda$, but their combinations cancel the
dependence. With the help of the optical theorem the inclusive decay
rate of a heavy quarkonium state $H$ could be expressed as:
\begin{equation}
\Gamma(H\rightarrow \mathrm{LH})=2\textrm{Im}\langle
H|\delta\mathcal{L}_{4-fermion}|H\rangle=\sum_{n}\frac{2\textrm{Im}f_{n}(\mu_\Lambda)}{m_{Q}^{d_{n}-4}}\langle
H|\mathcal{O}_{n}(\mu_\Lambda)|H\rangle.
\end{equation}
In principle, we need infinite terms to give theoretical
predictions, but in practice only a finite number of these terms are
needed to give an order of $v^2$ result, since the long-distance
matrix elements can be ordered in powers of $v^2$ by applying
the velocity scaling rules summarized in Ref.\cite{BBL2}. And the
short-distance coefficients (Wilson coefficients), defined in the matching
condition below, can be calculated perturbatively as a perturbation
series in QCD coupling constant $\alpha_{s}$
\begin{equation} \label{eqm}
\mathcal{A}(Q\overline{Q}\rightarrow
Q\overline{Q})\Big{|}_{\textrm{pert
QCD}}=\sum_{n}\frac{f_{n}(\mu_\Lambda)}{m^{d_{n}-4}}\langle
Q\overline{Q}|\mathcal{O}_{n}^{Q\overline{Q}}
(\mu_\Lambda)|Q\overline{Q}\rangle\Big{|}_{\textrm{pert NRQCD}}.
\end{equation}

When quark and antiquark are in a particular angular momentum state
$J$ and color state 1 or 8, the imaginary part of the left-hand side
of Eq.(\ref{eqm}) can be calculated with the covariant projection
method \cite{maltoni}:
\begin{align}
2\textrm{Im}\mathcal{A}((Q\overline{Q})_{^{2S+1}L_{J}}^{[1,8]}
\rightarrow(Q\overline{Q})_{^{2S+1}L_{J}}^{[1,8]})\Big{|}_{\textrm{pert
QCD}} \hspace{-4mm} =\frac{\langle
\mathcal{O}(^{2S+1}L_{J})\rangle_{QCD}}{K}\int\sum|\mathcal{\overline{M}}
((Q\overline{Q})_{^{2S+1}L_{J}}^{[1,8]}\rightarrow \mathrm{LH})|^{2}d\Phi,
\end{align}
where $\langle \mathcal{O}(^{2S+1}L_{J})\rangle_{QCD}$ equals to the
corresponding NRQCD four-fermion operator expectation value at tree
level, $L$ is the orbital angular momentum, $S$ is the total spin of the
heavy-quark pair, and $K$ is the degree of freedom of the initial
state. For spin-triplet states with $L$=0, $L=1$, and $L=2$, the relations
between $\mathcal{\overline{M}}(Q\overline{Q})_{^{3}L_{J}}^{[1,8]}$
and the full QCD Feynman amplitude $\mathcal{M}$ are
\begin{subequations}
\begin{align}
\mathcal{\overline{M}}((Q\overline{Q})_{^{3}S_{1}}^{[1,8]}\rightarrow
\mathrm{LH})=
\epsilon^{[S]}_{\rho}\mathrm{Tr}[\mathcal{C}^{[1,8]}\Pi^{\rho}\mathcal{M}]|_{q=0},
\end{align}
\begin{align}
\mathcal{\overline{M}}((Q\overline{Q})_{^{3}P_{P_{J}}}^{[8]}\rightarrow
\mathrm{LH})=\epsilon_{\alpha\rho}^{[P_{J}]}
\frac{\textrm{d}}{\textrm{d}q_{\alpha}}\mathrm{Tr}[\mathcal{C}^{[8]}\Pi^{\rho}\mathcal{M}]|_{q=0},
\end{align}
\begin{align}
\mathcal{\overline{M}}((Q\overline{Q})_{^{3}D_{D_{J}}}^{[1]}\rightarrow
\mathrm{LH})=\frac{1}{2}\epsilon_{\alpha\beta\rho}^{[D_{J}]}
\frac{\textrm{d}^2}{\textrm{d}q_{\alpha}\textrm{d}q_{\beta}}\mathrm{Tr}[\mathcal{C}^{[1]}\Pi^{\rho}\mathcal{M}]|_{q=0},
\end{align}
\end{subequations}
where $\epsilon^{s}_{\rho}, \epsilon^{P_{J}}_{\alpha\rho}$ and
$\epsilon^{D_{J}}_{\alpha\beta\rho}$ are the polarization tensors
for $L=S$, $P$, $D$ states with total angular momentum $1, P_{J}, D_{J}$,
respectively. For spin-triplet states, the spin projector of the
incoming heavy-quark pair accurate to all orders of $v^{2}$ is
\cite{v4}
\begin{equation}
\Pi^{\rho}=-\frac{1}{2\sqrt{2}(E+m_{Q})}(\frac{1}{2}\slashed{P}+\slashed{q}+m_{Q})\frac{\slashed{P}+2E}{2E}\gamma^{\rho}
(\frac{1}{2} \slashed{P}-\slashed{q}-m_{Q}),
\end{equation}
where $P^{\mu}$ is the four momentum of the heavy meson and
$P^{2}=4E^2$, and $2q^{\mu}$ is the relative momentum between the
quark and antiquark. The color projectors are
$\mathcal{C}^{[1]}=\frac{\delta_{i,j}}{\sqrt{N_c}}$ and
$\mathcal{C}^{[8]}=\sqrt{2}(T_{a})_{i,j}$.

In the Fock space, the $\psi(^3D_{J})$ states are represented by
\begin{equation}
|^3D_{J}\rangle=\mathcal{O}(1)|Q\bar{Q}(^3D_{J}^{[1]})\rangle+
\mathcal{O}(v)|Q\bar{Q}(^3P_{J}^{[8]})\rangle+
\mathcal{O}(v^{2})|Q\bar{Q}(^3S_{J}^{[1,8]})\rangle+\cdot\cdot\cdot.
\end{equation}
Here, the probability of $P$-wave and $S$-wave states are suppressed by
$v^2$ and $v^4$ relative to the $D$-wave state respectively, but their
operators scale as $v^{-2}$ and $v^{-4}$ relative to
$\mathcal{O}_{1}(^3D_{J})$. The relativistic effect and other Fock
state contributions are suppressed at least by $v^2$. Thus at
leading order in $v^2$ the NRQCD formula for $^{3}D_{J}$ decay into
LH is
\begin{eqnarray}\label{eq1}
&\Gamma&(^{3}D_{J}\rightarrow \mathrm{LH})=\nonumber\\
&2&\textrm{Im}f(^{3}D_{J}^{[1]})\frac{\langle
\psi(^{3}D_{J})|\mathcal{O}_{1}(^{3}D_{J})|\psi(^{3}D_{J})\rangle}{m_{Q}^{6}}
+\sum_{J=0}^{2}2\textrm{Im}f(^{3}P_{J}^{[8]})\frac{\langle
\psi(^{3}D_{J})|\mathcal{O}_{8}(^{3}P_{J_{P}})|\psi(^{3}D_{J})\rangle}{m_{Q}^{4}}+\nonumber\\
&2&\textrm{Im}f(^{3}S_{1}^{[8]})\frac{\langle \psi(^{3}D_{J})|
\mathcal{O}_{8}(^{3}S_{1})|\psi(^{3}D_{J})\rangle}{m_{Q}^{2}}+
2\textrm{Im}f(^{3}S_{1}^{[1]})\frac{\langle \psi(^{3}D_{J})|
\mathcal{O}_{1}(^{3}S_{1})|\psi(^{3}D_{J})\rangle}{m_{Q}^{2}},
\end{eqnarray}
where the four-fermion operators are defined\footnote{The
normalizations of the color singlet four-fermion operators agree
with those in Ref.\cite{maltoni}} as
\begin{subequations}
\begin{equation}
\mathcal{O}_{1}(^3S_{1})=\frac{1}{2N_{c}}\psi^{\dagger}\boldsymbol{\sigma}\chi
\cdot\chi^{\dagger}\boldsymbol{\sigma}\psi,
\end{equation}
\begin{equation}
\mathcal{O}_{8}(^3S_{1})=
\psi^{\dagger}\boldsymbol{\sigma}T^{a}\chi\cdot\chi^{\dagger}\boldsymbol{\sigma}T^{a}\psi,
\end{equation}
\begin{equation}
\mathcal{O}_{8}(^3P_{0})=\frac{1}{3}
\psi^{\dagger}(\frac{-i}{2}\overleftrightarrow{\boldsymbol{D}}\cdot\boldsymbol{\sigma})T^{a}\chi
\chi^{\dagger}(\frac{-i}{2}\overleftrightarrow{\boldsymbol{D}}\cdot\boldsymbol{\sigma})T^{a}\psi,
\end{equation}
\begin{equation}
\mathcal{O}_{8}(^3P_{1})=\frac{1}{2}
\psi^{\dagger}(\frac{-i}{2}\overleftrightarrow{\boldsymbol{D}}\times\boldsymbol{\sigma})T^{a}\chi
\cdot
\chi^{\dagger}(\frac{-i}{2}\overleftrightarrow{\boldsymbol{D}}\times\boldsymbol{\sigma})T^{a}\psi,
\end{equation}
\begin{equation}
\mathcal{O}_{8}(^3P_{2})=\frac{1}{2}
\psi^{\dagger}(\frac{-i}{2}\overleftrightarrow{\boldsymbol{D}}^{(i}\boldsymbol{\sigma}^{j)})T^{a}\chi~
\chi^{\dagger}(\frac{-i}{2}\overleftrightarrow{\boldsymbol{D}}^{(i}\boldsymbol{\sigma}^{j)})T^{a}\psi,
\end{equation}
\begin{equation}
\mathcal{O}_{1}(^{3}D_{1})=\frac{3}{10N_{c}}\psi^{\dagger}\boldsymbol{K}^{i}\chi~
\chi^{\dagger}\boldsymbol{K}^{i}\psi,
\end{equation}
\begin{equation}
\mathcal{O}_{1}(^{3}D_{2})=\frac{1}{12N_{c}}\psi^{\dagger}\boldsymbol{K}^{ij}\chi~
\chi^{\dagger}\boldsymbol{K}^{ij}\psi,
\end{equation}
\begin{equation}
\mathcal{O}_{1}(^{3}D_{3})=\frac{1}{18N_{c}}\psi^{\dagger}\boldsymbol{K}^{ijk}\chi~
\chi^{\dagger}\boldsymbol{K}^{ijk}\psi.
\end{equation}
\end{subequations}
The notations $\boldsymbol{K}$ are
$\boldsymbol{K}^{i}=\boldsymbol{\sigma}^{j}\boldsymbol{S}^{ij}$,
$\boldsymbol{K}^{ij}=\epsilon^{ikl}\boldsymbol{\sigma}^{l}\boldsymbol{S}^{jk}
+\epsilon^{jkl}\boldsymbol{\sigma}^{l}\boldsymbol{S}^{ik}$,
$\boldsymbol{K}^{ijk}=\boldsymbol{\sigma}^{i}\boldsymbol{S}^{jk}
+\boldsymbol{\sigma}^{j}\boldsymbol{S}^{ki}
+\boldsymbol{\sigma}^{k}\boldsymbol{S}^{ij}-\frac{2}{5}
\boldsymbol{\sigma}^{l}(\delta^{jk}\boldsymbol{S}^{il}
+\delta^{ki}\boldsymbol{S}^{jl} +\delta^{ij}\boldsymbol{S}^{kl})$,
where $\boldsymbol{S}^{ij}=(\frac{-i}{2})^{2}
(\overleftrightarrow{D}^{i}\overleftrightarrow{D}^{j}-\frac{1}{3}\overleftrightarrow{D}^{2}\delta^{ij})$.

For some processes, we need to calculate the NLO QCD corrections. To
handle the ultraviolet (UV) and infrared (IR) divergences in the
dimensional regularization scheme, one should extend the projection
method into $D=4-2\epsilon$ dimensions. The definitions of $\gamma$
matrixes in $D$ dimensions can be found in  quantum field theory
books. And the sums over polarization tensors
$\epsilon^{[S]}_{\rho}$, $\epsilon^{[P_{J}]}_{\alpha\rho}$ and
$\epsilon^{[D_{J}]}_{\alpha\beta\rho}$ in $D$ dimension are
\begin{subequations}
\begin{align}
\sum_{J_z}\epsilon^{(1)}_{\rho}\epsilon^{(1)\ast}_{\rho^{\prime}}=\Pi_{\rho
\rho^{\prime}},
\end{align}
\begin{align}
\sum_{J_z}\epsilon^{(0)}_{\alpha\rho}\epsilon^{(0)\ast}_{\alpha^{\prime}\rho^{\prime}}=
\frac{1}{D-1}\Pi_{\alpha\rho}\Pi_{\alpha^{\prime}\rho^{\prime}},
\end{align}
\begin{align}
\sum_{J_z}\epsilon^{(1)}_{\alpha\rho}\epsilon^{(1)\ast}_{\alpha^{\prime}\rho^{\prime}}=
\frac{1}{2}(\Pi_{\alpha\alpha^{\prime}}\Pi_{\rho\rho^{\prime}}
-\Pi_{\alpha\rho^{\prime}}\Pi_{\alpha^{\prime}\rho}),
\end{align}
\begin{align}
\sum_{J_z}\epsilon^{(2)}_{\alpha\rho}\epsilon^{(2)\ast}_{\alpha^{\prime}\rho^{\prime}}=
\frac{1}{2}(\Pi_{\alpha\alpha^{\prime}}\Pi_{\rho\rho^{\prime}}
+\Pi_{\alpha\rho^{\prime}}\Pi_{\alpha^{\prime}\rho})-
\frac{1}{D-1}\Pi_{\alpha\rho}\Pi_{\alpha^{\prime}\rho^{\prime}},
\end{align}
\begin{align}\label{3D1}
&\sum_{J_z}\epsilon_{\alpha\beta\rho}^{(1)}\epsilon_{\alpha'\beta'\rho
'}^{(1)*}=\frac{D-1}{2(D-2)(D+1)}(\Pi_{\alpha\rho}\Pi_{\alpha'\rho'}\Pi_{\beta\beta'}
+\Pi_{\beta\rho}\Pi_{\beta'\rho'}\Pi_{\alpha\alpha'}
+\Pi_{\alpha\rho}\Pi_{\beta'\rho'}\Pi_{\alpha'\beta}
+\Pi_{\beta\rho}\Pi_{\alpha'\rho'}\Pi_{\alpha\beta'}\nonumber\\
&-\frac{2}{D-1}(\Pi_{\alpha\rho}\Pi_{\alpha'\beta'}\Pi_{\beta\rho'}
+\Pi_{\beta\rho}\Pi_{\alpha'\beta'}\Pi_{\alpha\rho'}
+\Pi_{\alpha'\rho'}\Pi_{\alpha\beta}\Pi_{\beta'\rho}
+\Pi_{\beta'\rho'}\Pi_{\alpha\beta}\Pi_{\alpha'\rho})
+\frac{4}{(D-1)^{2}}\Pi_{\alpha\beta}\Pi_{\alpha'\beta'}\Pi_{\rho\rho'}),
\end{align}
\begin{align}\label{3D2}
\sum_{J_z}&\epsilon_{\alpha\beta\rho}^{(2)}\epsilon_{\alpha'\beta'\rho
'}^{(2)*}=\frac{1}{6}(2\Pi_{\alpha\alpha'}\Pi_{\beta\beta'}\Pi_{\rho\rho'}
+2\Pi_{\alpha\beta'}\Pi_{\alpha'\beta}\Pi_{\rho\rho'}
-\Pi_{\alpha\alpha'}\Pi_{\beta\rho'}\Pi_{\rho\beta'}
-\Pi_{\alpha\beta'}\Pi_{\beta\rho'}\Pi_{\rho\alpha'}
-\Pi_{\alpha\rho'}\Pi_{\beta\beta'}\Pi_{\rho\alpha'}\nonumber\\
&-\Pi_{\alpha\rho'}\Pi_{\beta\alpha'}\Pi_{\rho\beta'})+
\frac{1}{6(D-2)}(-4\Pi_{\alpha\beta}\Pi_{\alpha'\beta'}\Pi_{\rho\rho'}
+2\Pi_{\alpha\rho'}\Pi_{\alpha'\beta'}\Pi_{\beta\rho}
+2\Pi_{\alpha\rho}\Pi_{\beta\rho'}\Pi_{\alpha'\beta'}
+2\Pi_{\alpha\beta}\Pi_{\alpha'\rho}\Pi_{\beta'\rho'}\nonumber\\
&+2\Pi_{\alpha\beta}\Pi_{\beta'\rho}\Pi_{\alpha'\rho'}
-\Pi_{\alpha\beta'}\Pi_{\alpha'\rho'}\Pi_{\beta\rho}
-\Pi_{\alpha\rho}\Pi_{\alpha'\beta}\Pi_{\beta'\rho'}
-\Pi_{\alpha\alpha'}\Pi_{\beta\rho}\Pi_{\beta'\rho'}
-\Pi_{\alpha\rho}\Pi_{\beta\beta'}\Pi_{\alpha'\rho'}),
\end{align}
\begin{align}\label{3D3}
\sum_{J_z}&\epsilon_{\alpha\beta\rho}^{(3)}\epsilon_{\alpha'\beta'\rho
'}^{(3)*}=\frac{1}{6}(\Pi_{\alpha\alpha'}\Pi_{\beta\beta'}\Pi_{\rho\rho'}
+\Pi_{\alpha\alpha'}\Pi_{\beta\rho'}\Pi_{\rho\beta'}
+\Pi_{\alpha\beta'}\Pi_{\beta\alpha'}\Pi_{\rho\rho'}
+\Pi_{\alpha\beta'}\Pi_{\beta\rho'}\Pi_{\rho\alpha'}
+\Pi_{\alpha\rho'}\Pi_{\beta\beta'}\Pi_{\rho\alpha'}\nonumber\\
&+\Pi_{\alpha\rho'}\Pi_{\beta\alpha'}\Pi_{\rho\beta'})-\frac{1}{3(D+1)}
(\Pi_{\alpha\beta}\Pi_{\rho\alpha'}\Pi_{\beta'\rho'}
+\Pi_{\alpha\beta}\Pi_{\rho\beta'}\Pi_{\alpha'\rho'}
+\Pi_{\alpha\beta}\Pi_{\rho\rho'}\Pi_{\alpha'\beta'}
+\Pi_{\alpha\rho}\Pi_{\beta\alpha'}\Pi_{\beta'\rho'}\nonumber\\
&+\Pi_{\alpha\rho}\Pi_{\beta\beta'}\Pi_{\alpha'\rho'}
+\Pi_{\alpha\rho}\Pi_{\beta\rho'}\Pi_{\alpha'\beta'}
+\Pi_{\beta\rho}\Pi_{\alpha\alpha'}\Pi_{\beta'\rho'}
+\Pi_{\beta\rho}\Pi_{\alpha\beta'}\Pi_{\alpha'\rho'}
+\Pi_{\beta\rho}\Pi_{\alpha\rho'}\Pi_{\alpha'\beta'}).
\end{align}
\end{subequations}
And the degrees of freedom are $D-1$ for the $S$-wave state;  1,
$\frac{(D-1)(D-2)}{2}$, $\frac{(D+1)(D-2)}{2}$ for $J$=0, 1, 2 $P$-wave
states; and  $D-1$, $\frac{(D-3)(D-1)(D+1)}{3}$,
$\frac{(D-2)(D-1)(D+3)}{6}$ for $J=1$, 2, 3 $D$-wave states. The rather
trivial extensions for $L=0$ and 1 cases were given in
Ref.\cite{maltoni}. Here three principles are adopted to construct
the nontrivial results for the $D$-wave case. First, the symmetry of
the three indexes should be kept; second, the inner products between
one tensor and the other two are zero; third, the completeness
condition should be satisfied:
\begin{eqnarray}\label{complete}
\sum_{J_z}\epsilon_{\alpha\beta\rho}^{(1)}\epsilon_{\alpha'\beta'\rho
'}^{(1)*}+\sum_{J_z}\epsilon_{\alpha\beta\rho}^{(2)}\epsilon_{\alpha'\beta'\rho
'}^{(2)*}+\sum_{J_z}\epsilon_{\alpha\beta\rho}^{(3)}\epsilon_{\alpha'\beta'\rho
'}^{(3)*}=\nonumber\\
(\frac{1}{2}(\Pi_{\alpha\alpha^{\prime}}\Pi_{\beta\beta^{\prime}}
+\Pi_{\alpha\beta^{\prime}}\Pi_{\alpha^{\prime}\beta})-
\frac{1}{D-1}\Pi_{\alpha\beta}\Pi_{\alpha^{\prime}\beta^{\prime}})\Pi_{\rho\rho^{\prime}}.
\end{eqnarray}
We calculate the degrees of freedom of each $^3D_{J}$ state with
group theory and get Eq.(\ref{3D1}). Then we derive out
Eq.(\ref{3D3}) with the help of the first two principles. In the end
Eq.(\ref{3D2})=Eq.(\ref{complete})-Eq.(\ref{3D1})-Eq.(\ref{3D3}).

To get the NLO NRQCD results, the operator mixing equation (\ref{CJPD})
between $P$-wave and $D$-wave operators in momentum space should also be
extended into $D$ dimension for consistency:
\begin{equation}\label{CJPD}
\sum_{D_{J'}}C_{P_{J},D_{J'}}\langle(Q\bar{Q})_{^3D_{J'}}|\mathcal{O}(^3D_{J'})|(Q\bar{Q})_{^3D_{J'}}\rangle=
\langle(Q\bar{Q})_{^3P_{J}}|\mathcal{O}(^3P_{J})|(Q\bar{Q})_{^3P_{J}}\rangle
\boldsymbol{\vec{q}}\cdot \boldsymbol{\vec{q}^{\prime}},
\end{equation}
where $C_{P_{J},D_{J'}}$ are the generalized Clebsch-Gordan
coefficients:
\begin{equation}\label{CJ}
C_{P_{J},D_{J}}=\frac{|\epsilon^{\alpha\beta\rho(D_{J})}\epsilon^{\ast
(P_{J})}_{\alpha\rho}\epsilon^{\ast
(S=1)}_{\beta}|^{2}}{\epsilon^{(D_{J})}_{\alpha^{\prime}\beta^{\prime}\rho^{\prime}}
\epsilon^{\ast\alpha^{\prime}\beta^{\prime}\rho^{\prime}(D_{J})}},
\end{equation}
where the repeated indexes mean being summed in $D$ dimension.

After resolving the above problems, one can do calculations from
both full QCD and NRQCD straightforwardly.

\section{Full QCD Calculation}
In Sec.II, it has been explained that at leading order of
$v^{2}$ the LH decays of $^3D_{J}$ contain the subprocesses of
$Q\bar{Q}_{^3S_{1}}^{[1,8]},$ $Q\bar{Q}_{^3P_{J}}^{[8]}(J=0,1,2)$,
and $Q\bar{Q}_{^{3}D_{J}}^{[1]}(J=1,2,3)$ annihilating into gluons
or light quarks. When doing the calculation with full QCD theory,
for simplicity, only the explicit imaginary part of
$\mathcal{\bar{A}}$ is given here,
\begin{equation}
2\textrm{Im}\mathcal{\bar{A}}((Q\overline{Q})_{^{2S+1}L_{J}}^{[1,8]}
\rightarrow(Q\overline{Q})_{^{2S+1}L_{J}}^{[1,8]})\Big{|}_{\textrm{pert
QCD}}=\frac{1}{K}\int\sum|\mathcal{\overline{M}}((Q\overline{Q})_{^{2S+1}L_{J}}^{[1,8]}\rightarrow
\mathrm{LH})|^{2}d\Phi.
\end{equation}
In the following subsections, the contributions at $\alpha_{s}^{2}$
order, and the corresponding real and virtual corrections will be
given in Secs.III\textbf{A}, III\textbf{B} and III\textbf{C}
respectively. As for the processes whose tree level diagrams are
already at $\alpha_{s}^{3}$ order, their results will also be given
in Sec.III.

\subsection{LO Results}

There are three subprocesses $(Q\bar{Q})_{^3S_{1}}^{[8]}\rightarrow
q\bar{q}$, $(Q\bar{Q})_{^3P_{0,2}}^{[8]}\rightarrow gg$ with
nonvanishing imaginary parts at $\mathcal{O}(\alpha^{2}_{s})$.  The
Feynman diagrams are shown in Fig.[1]. And the results in
$D$ dimension are
\begin{subequations}
\begin{equation}
(2\textrm{Im}\mathcal{\bar{A}}(^3S_{1}^{[8]}))^{\textrm{Born}}=
\frac{1-\epsilon}{3-2\epsilon}\frac{8N_{f}\alpha_{s}^{2}\pi^{2}\mu^{4\epsilon}
\Phi_{(2)}}{m_{Q}^{2}},
\end{equation}
\begin{equation}
(2\textrm{Im}\mathcal{\bar{A}}(^3P_{0}^{[8]}))^{\textrm{Born}}=
\frac{1-\epsilon}{3-2\epsilon}\frac{144B_{F}\alpha_{s}^{2}\pi^2\mu^{4\epsilon}
\Phi_{(2)}}{m_{Q}^{4}},
\end{equation}
\begin{equation}
(2\textrm{Im}\mathcal{\bar{A}}(^3P_{2}^{[8]}))^{\textrm{Born}}=
\frac{4\epsilon^2-13\epsilon+6}{(3-2\epsilon)(5-2\epsilon)}\frac{32B_{F}\alpha_{s}^{2}\pi^2\mu^{4\epsilon}
\Phi_{(2)}}{m_{Q}^{4}},
\end{equation}
\end{subequations}
where
$\Phi_{(2)}=\frac{1}{8\pi}\frac{\Gamma(1-\epsilon)}{\Gamma(2-2\epsilon)}(\frac{\pi}{m_{Q}^{2}})^{\epsilon}$
is the two-body phase space in $D$ dimension.

\begin{figure}
\begin{center}
\includegraphics[scale=0.7]{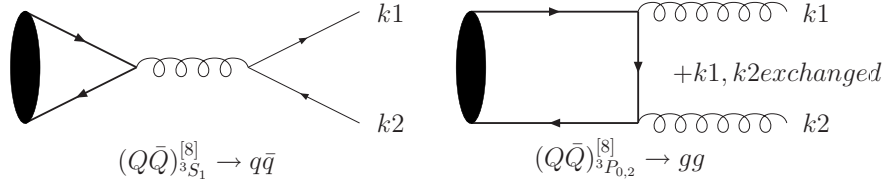}
\caption{Feynman diagrams for$(Q\bar{Q})_{^3S_{1}}^{[8]}\rightarrow
q\bar{q}$, $(Q\bar{Q})_{^3P_{0,2}}^{[8]}\rightarrow gg$}
\end{center}
\end{figure}

\subsection{Real Corrections}
Besides the real corrections (RC), other processes with three bodies
in final states include $(Q\bar{Q})_{^3S_{1}}^{[1,8]}\rightarrow3g$,
$(Q\bar{Q})_{^3P_{1}}^{[8]}\rightarrow3g$,$(Q\bar{Q})_{^3P_{J}}^{[8]}\rightarrow
q\bar{q}g$ and $(Q\bar{Q})_{^3D_{J}}^{[1]}\rightarrow 3g$. And the
typical Feynman diagrams for $q\bar{q}g$ and $3g$ are shown in
Fig.[2] and Fig.[3], respectively.\footnote{ In Feynman gauge the
ghost diagrams are also needed, where three-gluon or four-gluon
vertex will appear.}
\begin{figure}
\begin{center}
\includegraphics[scale=0.7]{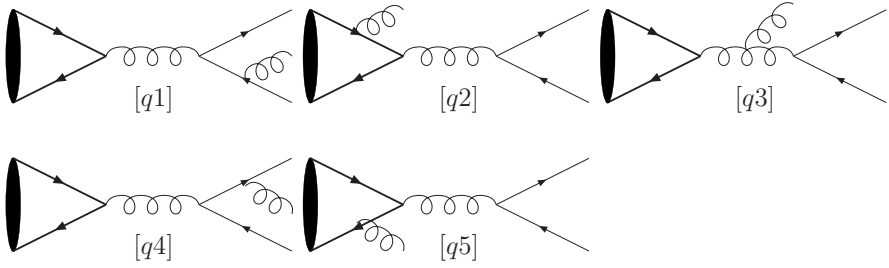}
\caption{ Feynman diagrams for
$(Q\bar{Q})_{^{3}L_{J}}^{[8]}\rightarrow q\bar{q}g$ }
\end{center}
\end{figure}

\begin{figure}
\begin{center}
\includegraphics[scale=0.7]{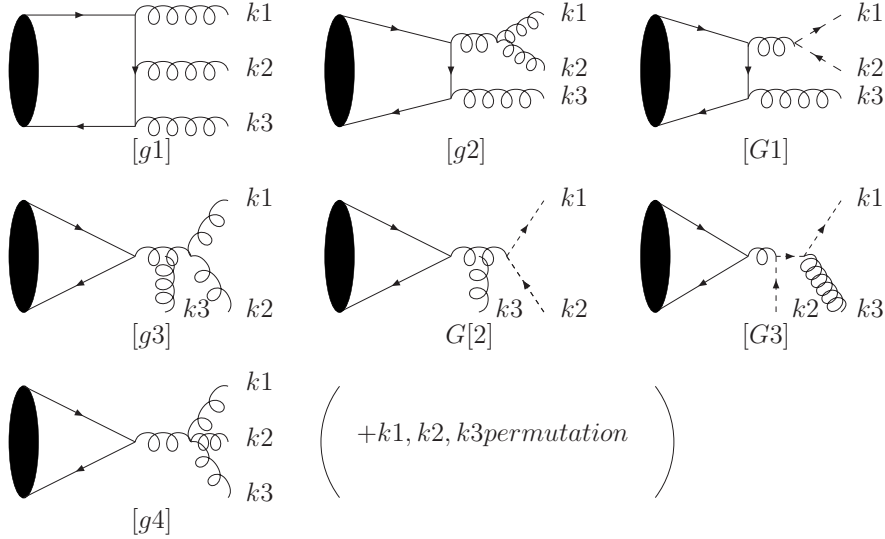}
\caption{ Feynman diagrams for
$(Q\bar{Q})_{^{3}L_{J}^{[1,8]}}\rightarrow 3g$. Diagrams with
different positions of $k_{i}$ are neglected. In Feynman gauge the
ghost diagrams should also be contained. }
\end{center}
\end{figure}

We calculate the Feynman amplitudes and do the phase space
integrations both in $D$ dimension, and check the results with eikonal
approximation relations or Altarelli-Parisi splitting functions when
one gluon is soft or two light partons (gluon or light quark) are
collinear. Three variables $x_{i}=\frac{E_{i}}{m_{Q}}$ are
introduced to describe the $(Q\bar{Q})_{^3L_{J}^{[1,8]}}\rightarrow
k_{1}+k_{2}+k_{3}$ process. For energy conservation
$x_{1}+x_{2}+x_{3}=2$. The three-body phase space in $D$ dimension
becomes
\begin{equation}
d\Phi_{(3)}=\Phi_{(2)}\frac{4m_{Q}^{2}}{(4\pi)^{2}\Gamma(1-\epsilon)}(\frac{\pi}{m_{Q}^{2}})^{^\epsilon}
[(1-x_{1})(1-{x_2})(1-x_{3})]^{^{-\epsilon}}\delta(2-x_{1}-x_{2}-x_{3})dx_{1}dx_{2}dx_{3}.
\end{equation}
In the phase space, the $x_{i}=0$ region is the soft region of
particle with momentum $k_{i}$, and the $x_{i}=1$ region is the
collinear region of the other two particles. And the inner products
of those four momenta are $P\cdot k_{i}=2m_{Q}^{2}x_{i}$, $
k_{i}\cdot k_{j}=2m_{Q}^{2}(x_{i}+x_{j}-1)$.

When $L=S$ and $P$, the real corrections contributing to
$2\textrm{Im}\mathcal{\bar{A}}(^3L_{J}^{[1,8]})^{^{\mathrm{RC}}}$ are
\begin{eqnarray}
2\textrm{Im}\mathcal{\bar{A}}(^3L_{J}^{[1,8]})^{^{\mathrm{RC}}}
=&\frac{1}{3!(K_{c1,c8})(K_{J})}\int\sum|\mathcal{\overline{M}}((Q\overline{Q})_{^{3}L_{J}}^{[1,8]}\rightarrow
ggg)|^{2}d\Phi_{(3)}\nonumber\\
&+\frac{1}{(K_{c1,c8})(K_{J})}\int\sum|\mathcal{\overline{M}}((Q\overline{Q})_{^{3}L_{J}}^{[1,8]}\rightarrow
q\bar{q}g)|^{2}d\Phi_{(3)},
\end{eqnarray}
where $K_{c1}=1$, $K_{c8}=N_{c}^2-1$, and $K_{J}$ is the
polarization number of angular momentum $J$ state. In the $D$-wave case,
because of parity conservation, only the $3g$ process is left.

\subsubsection{$(Q\bar{Q})_{^3L_{J}^{[1,8]}}\rightarrow ggg $ }
There are 18 diagrams in the $Q\bar{Q}\to ggg$ process. Because of
$J^{PC}$ conservation,  only class [g1] is needed in $^3S_{1}^{[1]}
\rightarrow ggg$ and $^3D_{J}^{[1]}\rightarrow ggg$ processes. In
the $^3P_{J}^{[8]}\rightarrow ggg$ processes, the diagrams are those in
class [g1], class [g2] and the corresponding ghost ones in class [G1].
And all the diagrams should be calculated in the
$^3S_{1}^{[8]}\rightarrow ggg$ case. We now proceed to show how to
get the physical results at a differential cross section level when
ghost diagrams are contained. In the Feynman gauge, when the three or four
gluon vertex appears, the nonphysical (NP) degrees of freedom  are
removed by ghost diagrams. Label the square of the amplitudes of
the gluon diagrams in the Feynman gauge with the subscript NP, and express the ghost result as
\begin{equation}
\mathcal|{\overline{M}}_{G}|^{2}_{k_{i},k_{j}}=\sum|\mathcal{\overline{M}}((Q\overline{Q})_{^{3}L_{J}}^{[1,8]}\rightarrow
gG_{k_{i}}\bar{G}_{k_{j}})|^{2},
\end{equation}
where $G_{k_{i}}$ and $\bar{G}_{k_{j}}$ are for ghost and antighost
with momentum $k_{i}$ and $k_{j}$, respectively. To cancel the NP
part of each gluon, all possibilities of indexes  $i,j$ should be
summed over. Then proper physical results at the differential level are
\begin{eqnarray}
\sum|\mathcal{\overline{M}}((Q\overline{Q})_{^{3}L_{J}}^{[1,8]}\rightarrow
ggg)|^{2}=\sum|\mathcal{\overline{M}}_{NP}((Q\overline{Q})_{^{3}L_{J}}^{[1,8]}\rightarrow
ggg)|^{2}-\sum_{i,j}\mathcal|{\overline{M}}_{G}|^{2}_{k_{i},k_{j}}
\end{eqnarray}

Here we omit the details and only give the real corrections below
\begin{subequations}\label{FullQCDSPD}
\begin{align}
\frac{1}{3!(3-2\epsilon)}\int\sum|\mathcal{\overline{M}}((Q\overline{Q})_{^{3}S_{1}}^{[1]}\rightarrow
ggg)|^{2}d\Phi_{(3)}=
\frac{40\alpha_{s}^{3}(\pi^{2}-9)}{81m_{Q}^{2}},
\end{align}
\begin{align}
\frac{1}{3!(N_{C}^2-1)(3-2\epsilon)}\int\sum|\mathcal{\overline{M}}((Q\overline{Q})_{^{3}S_{1}}^{[8]}\rightarrow
ggg)|^{2}d\Phi_{(3)}=
\frac{5(67\pi^2-657)\alpha_{s}^{3}}{108m_{Q}^{2}},
\end{align}
\begin{align}
\frac{1}{3!(N_{C}^{2}-1)}\int\sum|\mathcal{\overline{M}}((Q\overline{Q})_{^{3}P_{0}}^{[8]}\rightarrow
ggg)|^{2}d\Phi_{(3)}=\nonumber\\
\frac{C_{A}\alpha_{s}F_{\epsilon}}{\pi}
(\frac{1}{\epsilon^{2}}+\frac{7}{3\epsilon}+\frac{875-60\pi^2}{162})
(2\textrm{Im}\mathcal{\bar{A}}(^3P_{0}^{[8]}))^{\textrm{Born}},
\end{align}
\begin{align}
\frac{2}{3!(N_{C}^{2}-1)(3-2\epsilon)(2-2\epsilon)}\int\sum|\mathcal{\overline{M}}((Q\overline{Q})_{^{3}P_{1}}^{[8]}\rightarrow
ggg)|^{2}d\Phi_{(3)}=
\frac{C_{A}B_{F}(-138\pi^2+1369)\alpha_{s}^{3}}{54m_{Q}^{4}},
\end{align}
\begin{align}
\frac{2}{3!(N_{C}^{2}-1)(5-2\epsilon)(2-2\epsilon)}\int\sum|\mathcal{\overline{M}}((Q\overline{Q})_{^{3}P_{2}}^{[8]}\rightarrow
ggg)|^{2}d\Phi_{(3)}=\nonumber\\
\frac{C_{A}\alpha_{s}F_{\epsilon}}{\pi}
(\frac{1}{\epsilon^{2}}+\frac{7}{3\epsilon}+\frac{4679-438\pi^2}{432})
(2\textrm{Im}\mathcal{\bar{A}}(^3P_{2}^{[8]}))^{\textrm{Born}},
\end{align}
\begin{align}
&\frac{1}{3!(3-2\epsilon)}\int\sum|\mathcal{\overline{M}}((Q\overline{Q})_{^{3}D_{1}}^{[1]}\rightarrow
ggg)|^{2}d\Phi_{(3)}
=\nonumber\\
&\frac{32}{3m_{Q}^{6}}B_{F}\Phi_{2}F_{\epsilon}\pi\alpha_{s}^{3}\mu^{4\epsilon}
(-\frac{608}{135\epsilon}+\frac{-7744+1605\pi^2}{16200})
\end{align}
\begin{align}
&\frac{3}{3!(5-2\epsilon)(3-2\epsilon)(1-2\epsilon)}\int\sum|\mathcal{\overline{M}}((Q\overline{Q})_{^{3}D_{2}}^{[1]}\rightarrow
ggg)|^{2}d\Phi_{(3)}=\nonumber\\
&\frac{32}{3m_{Q}^{6}}B_{F}\Phi_{2}F_{\epsilon}\pi\alpha_{s}^{3}\mu^{4\epsilon}
(-\frac{8}{15\epsilon}+\frac{-23024+2125\pi^2}{1800})
\end{align}
\begin{align}
&\frac{6}{3!(7-2\epsilon)(3-2\epsilon)(2-2\epsilon)}\int\sum|\mathcal{\overline{M}}((Q\overline{Q})_{^{3}D_{3}}^{[1]}\rightarrow
ggg)|^{2}d\Phi_{(3)}=\nonumber\\
&\frac{32}{3m_{Q}^{6}}B_{F}\Phi_{2}F_{\epsilon}\pi\alpha_{s}^{3}\mu^{4\epsilon}
(-\frac{32}{15\epsilon}+\frac{-28656+2645\pi^{2}}{6300})
\end{align}
\end{subequations}
where the "=" are correct at $\mathcal{O}(1)$,
$F_{\epsilon}=(\frac{\pi\mu^{2}}{m_{Q}^{2}})^{^{\epsilon}}\Gamma(1+\epsilon)$
,$C_{A}=3$, and $B_{F}=5/12$ are color factors. The  expressions of
$\sum|\overline{\mathcal{M}}((Q\overline{Q})_{^{3}D_{J}}^{[1]}\rightarrow
ggg)|^{2}$ are too complicated to be given here.

\subsubsection{$(Q\bar{Q})_{^3L_{J}^{[1,8]}}\rightarrow q\bar{q}g$}

At $\mathcal{O}(\alpha_{s}^{3})$ there are four subprocesses:
$(Q\bar{Q})_{^3S_{1}^{[8]}}\rightarrow q\bar{q}g$, which include all
the Feynman diagrams in Fig.[2], and
$(Q\bar{Q})_{^3P_{J}^{[8]}}\rightarrow q\bar{q}g$, in which there
are only two diagrams $q[2]$ and $q[5]$. As in the $3g$ processes
both the Feynman amplitudes and the phase space integrals are
calculated in $D$ dimension directly and the results are
\begin{subequations}\label{qqg}
\begin{eqnarray}
&\frac{1}{(N_{c}^2-1)(3-2\epsilon)}\int\sum|\mathcal{\overline{M}}((Q\overline{Q})_{^{3}S_{1}}^{[8]}\rightarrow
q\bar{q}g)|^{2}d\Phi_{(3)}=&\nonumber\\
&\frac{\alpha_{s}F_{\epsilon}}{\pi}
\Big{(}C_{F}(\frac{1}{\epsilon^2}+\frac{3}{2\epsilon}+\frac{57-8\pi^2}{12})+C_{A}(\frac{1}{2\epsilon}+\frac{11}{6})\Big{)}
(2\textrm{Im}\mathcal{\bar{A}}(^3S_{1}^{[8]}))^{\textrm{Born}}&
\end{eqnarray}
\begin{eqnarray}
&\frac{1}{(N_{c}^2-1)}\int\sum|\mathcal{\overline{M}}((Q\overline{Q})_{^{3}P_{0}}^{[8]}\rightarrow
q\bar{q}g)|^{2}d\Phi_{(3)}=&\nonumber\\
&(N_f\frac{\alpha_{s}F_{\epsilon}}{\pi}(-\frac{1}{3\epsilon}))(2\textrm{Im}\mathcal{\bar{A}}(^3P_{0}^{[8]}))^{\textrm{Born}}
-\frac{4}{9m_{Q}^{2}}B_{F}\alpha_{s}
(3(\textrm{2Im}\mathcal{\bar{A}}(^3S_{1}^{[8]}))^{\textrm{Born}}\frac{F_{\epsilon}}{\pi\epsilon}
+\frac{29N_{f}}{3m_{Q}^{2}}\alpha_{s}^{2})&
\end{eqnarray}
\begin{eqnarray}
&\frac{1}{(N_{c}^2-1)}\int\sum|\mathcal{\overline{M}}((Q\overline{Q})_{^{3}P_{1}}^{[8]}\rightarrow
q\bar{q}g)|^{2}d\Phi_{(3)}=&\nonumber\\
&(N_f\frac{\alpha_{s}F_{\epsilon}}{\pi}(-\frac{1}{3\epsilon}))(2\textrm{Im}\mathcal{\bar{A}}(^3P_{1}^{[8]}))^{\textrm{Born}}
-\frac{4}{9m_{Q}^{2}}B_{F}\alpha_{s}
(3(\textrm{2Im}\mathcal{\bar{A}}(^3S_{1}^{[8]}))^{\textrm{Born}}\frac{F_{\epsilon}}{\pi\epsilon}
+\frac{8N_{f}}{3m_{Q}^{2}}\alpha_{s}^{2})&
\end{eqnarray}
\begin{eqnarray}
&\frac{1}{(N_{c}^2-1)}\int\sum|\mathcal{\overline{M}}((Q\overline{Q})_{^{3}P_{2}}^{[8]}\rightarrow
q\bar{q}g)|^{2}d\Phi_{(3)}=&\nonumber\\
&(N_f\frac{\alpha_{s}F_{\epsilon}}{\pi}(-\frac{1}{3\epsilon}))(2\textrm{Im}\mathcal{\bar{A}}(^3P_{2}^{[8]}))^{\textrm{Born}}
-\frac{4}{9m_{Q}^{2}}B_{F}\alpha_{s}
(3(\textrm{2Im}\mathcal{\bar{A}}(^3S_{1}^{[8]}))
^{\textrm{Born}}\frac{F_{\epsilon}}{\pi\epsilon}
+\frac{58N_{f}}{15m_{Q}^{2}}\alpha_{s}^{2})&
\end{eqnarray}
\end{subequations}

\subsection{Virtual Corrections}
The virtual corrections are performed with a renormalized Lagrangian,
in which the renormalization constants of the QCD gauge coupling
constant $g_{s}=\sqrt{4\pi\alpha_{s}}$, heavy-quark $m_{Q}$,
heavy-quark field $\psi_{Q}$, light quark field $\psi_{q}$, and gluon field
$\mathcal{A}_{\mu}$ are defined as
\begin{equation}
g_{s}^{0}=Z_{g}g_{s},\quad m_{Q}^{0}=Z_{m_{Q}}m_{Q},\quad
\psi_{Q}^{0}=\sqrt{Z_{2Q}}\psi,\quad
\psi_{q}^{0}=\sqrt{Z_{2q}}\psi,\quad
\mathbf{A}_{\mu}^{0}=\sqrt{Z_{3}}\mathcal{A}_{\mu},
\end{equation}
where the superscript 0 labels bare quantities, and
$Z_{i}=1+\delta_{i}$. Here a mixing renormalization scheme
\cite{Kniehl} is adopted. The quark mass $m_{Q}$, heavy-quark field
$\psi_{Q}$, light quark field $\psi_{q}$, and gluon field
$\mathbf{A}_{\mu}$ are defined in the on-shell condition, while
$g_{s}$ is  in the minimal-subtraction($\overline{MS}$) scheme. Then
in this mixing scheme, these renormalization constants are
\begin{subequations}
\begin{equation}
\delta Z_{2Q}^{OS}=-\frac{C_{F}\alpha_{s}F_{\epsilon}}{4\pi}
(\frac{1}{\epsilon_{_{UV}}}+\frac{2}{\epsilon_{_{IR}}}+3\ln(4)+4+\mathcal{O}(\epsilon))
\end{equation}
\begin{equation}
\delta Z_{2q}^{OS}=-\frac{C_{F}\alpha_{s}F_{\epsilon}}{4\pi}
(\frac{1}{\epsilon_{_{UV}}}-\frac{1}{\epsilon_{_{IR}}}+\mathcal{O}(\epsilon))
\end{equation}
\begin{equation}
\delta Z_{3}^{OS}=\frac{\alpha_{s}F_{\epsilon}}{4\pi}
(\beta_{0}-C_{A})(\frac{2}{\epsilon_{_{UV}}}-\frac{2}{\epsilon_{_{IR}}}+\mathcal{O}(\epsilon))
\end{equation}
\begin{equation}
\delta Z_{m_{Q}}^{OS}=-\frac{3C_{F}\alpha_{s}F_{\epsilon}}{4\pi}
(\frac{1}{\epsilon_{_{UV}}}+\ln(4)+\frac{4}{3}+\mathcal{O}(\epsilon))
\end{equation}
\begin{equation}
\delta
Z_{g}^{\overline{MS}}=-\frac{\beta_{0}\alpha_{s}F_{\epsilon}}{4\pi}
(\frac{1}{\epsilon_{_{UV}}}-\ln(\frac{\mu^2}{4m_{Q}^{2}})+\mathcal{O}(\epsilon))
\end{equation}
\end{subequations}
where $\beta_{0}=\frac{11C_{A}}{6}-\frac{N_{f}}{3}$, and $N_{f}$ is
the number of light flavor quarks. The representative virtual correction Feynman
diagrams for the Born processes in Sec.III$\textbf{B}$ are shown in
Figs.4 and 5, without external leg correction diagrams in
this scheme. The UV divergences in self-energy and triangle diagrams
will be canceled by the counterterm diagrams correspondingly.
\begin{figure}
\begin{center}
\includegraphics[scale=0.65]{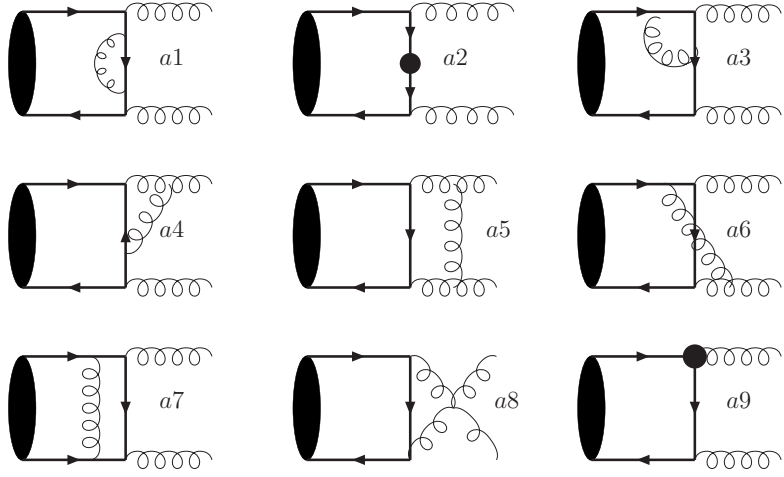}
\caption{One-loop Feynman diagram for
$(Q\bar{Q})_{^{3}P_{0,2}^{[8]}}\rightarrow gg$ }
\end{center}
\end{figure}
To regularize the Coulomb $\frac{1}{v}$ poles in the virtual
processes, the loop integrals are done first before setting the
relative momentum $q=0$. Also in
$(Q\bar{Q})_{^3P_{0,2}}^{[8]}\rightarrow gg$ processes, we integrate
the loop momentum first then compute the first derivative of the
Feynman amplitudes with respect to $q^{\alpha}$. When $q\neq0$,  the
momenta of $Q$ and $\bar{Q}$ are $P_{Q}=\frac{P}{2}+q$,
$P_{\bar{Q}}=\frac{P}{2}-q$, where $P$ is the meson total momentum,
and the momenta of the two massless final state particles are
labeled with $k_{1}$ and $k_{2}$. Then the scalar functions of the
loop integrals can be expressed in the Mandelstam variables:
\begin{equation}
s=(P_{Q}+P_{\bar{Q}})^2,\quad t=(P_{Q}-k_{1})^2, \quad
u=(P_{Q}-k_{2})^2,
\end{equation}
where $s=4m_{Q}^2/(1-v^2)$. We do the calculations diagram by
diagram and summarize the results in the following form:
\begin{equation}
2\textrm{Im}\mathcal{\bar{A}}(^3L_{J}^{[1,8]})^{^{VC}}=
2\textrm{Im}\mathcal{\bar{A}}(^3L_{J}^{[1,8]})^{\textrm{Born}}\frac{\alpha_{s}F_{\epsilon}}{\pi}
\sum_{k}D_{k},
\end{equation}
where $D_{k}$ for each process are listed in Table[I], [II], [III].
The contributions of the counter-term diagrams are put together with
the corresponding self-energy and vertex diagrams to show explicit
cancelation of the UV divergences. The IR divergences left will be
canceled by the RC.

\begin{figure}
\begin{center}
\includegraphics[scale=0.65]{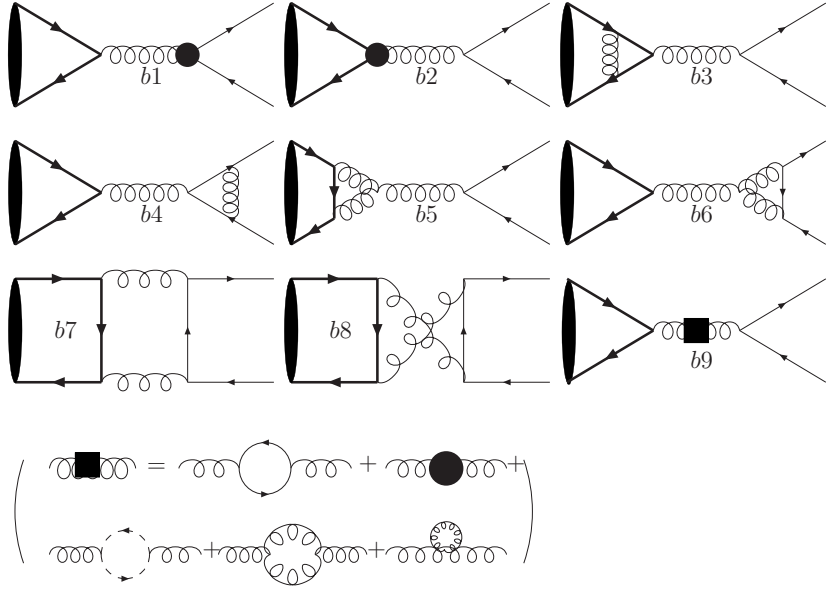}
\caption{One-loop Feynman diagram for
$(Q\bar{Q})_{^{3}S_{1}^{[8]}}\rightarrow q\bar{q}$ }
\end{center}
\end{figure}

\begin{center}
\begin{table}
\caption{Virtual corrections to
$(Q\bar{Q})_{^3S_{1}^{[8]}}\rightarrow q\bar{q}$.}
\begin{tabular}{|c|c|}
     \hline
     Diag.& $D_{k}$\\
     \hline
      $b1+b4+b6$ & $\frac{C_{A}-2C_{F}}{2\epsilon_{_{IR}}^2}+\frac{C_{A}-3C_{F}-\beta_{0}}{2\epsilon_{_{IR}}}+
      \frac{C_{A}(9-2\pi^2)+4C_{F}(-6+\pi^2)}{6}+\frac{\beta_{0}}{2}\ln(\frac{\mu^{2}}{4m_{Q}^{2}})$ \\
     \hline
      $b2+b3+b5$ & $\frac{(2C_{F}-C_{A})\pi^2}{4v}-\frac{\beta_{0}}{2\epsilon_{_{IR}}}+
      \frac{C_{A}(7+2\ln(2))-12C_{F}}{3}+\frac{\beta_{0}}{2}\ln(\frac{\mu^{2}}{4m_{Q}^{2}})$ \\
      \hline
     $b7$ & $(2C_{F}-\frac{C_{A}}{2})(\frac{\pi^2}{6}-\frac{1}{\epsilon_{_{IR}}^{2}})$ \\
     \hline
     $b8$ & $(2C_{F}-C_{A})(\frac{1}{\epsilon_{_{IR}}^{2}}-\frac{\pi^2}{6})$\\
     \hline
     $b9$ &
     $(\frac{5}{6}C_{A}-\frac{1}{3}N_{f})\frac{1}{\epsilon_{_{IR}}}+\frac{31}{18}C_{A}-\frac{5}{9}N_{f}$\\
     \hline
\end{tabular}
\end{table}
\end{center}

\begin{center}
\begin{table}
\caption{Virtual corrections to
$(Q\bar{Q})_{^3P_{0}^{[8]}}\rightarrow gg$.}
\begin{tabular}{|c|c|}
     \hline
     Diag.& $D_{k}$\\
     \hline
       $a1+a2$ & $\frac{C_{F}}{\epsilon_{_{IR}}}+\frac{C_{F}(5+32\ln(2))}{9}$\\
     \hline
     $a3+a4+a9$  &$\frac{-C_{A}}{3\epsilon_{_{IR}}^{2}}+\frac{-9\beta_{0}-C_{A}-18C_{F}}{9\epsilon_{_{IR}}}+
     \frac{C_{F}(-44-128\ln(2)+3\pi^2)}{18}+\frac{C_{A}(38+8\ln(2)-\pi^2)}{36}
     +\beta_{0}\ln(\frac{\mu^{2}}{4m_{Q}^{2}})$ \\
     \hline
     $a5$  & $\frac{C_{A}}{2}(\frac{-4}{3\epsilon^{2}_{_{IR}}}+\frac{2}{9\epsilon_{_{IR}}}-
     \frac{14}{27}+\frac{13}{18}\pi^2+\frac{14}{27}\ln(2))$ \\
     \hline
     $a6$  & 0 \\
      \hline
     $a7$  & $(C_{F}-\frac{C_{A}}{2})(\frac{1}{\epsilon_{_{IR}}}+\frac{\pi^2}{2v}-
     \frac{4}{9}+\frac{1}{12}\pi^2+\frac{32}{9}\ln(2))$ \\
      \hline
     $a8$  & $C_{A}(\frac{-19}{27}+\frac{70}{27}\ln(2))$ \\
      \hline
\end{tabular}
\end{table}
\end{center}

\begin{center}
\begin{table}
\caption{Virtual corrections to
$(Q\bar{Q})_{^3P_{2}^{[8]}}\rightarrow gg$.}
\begin{tabular}{|c|c|}
     \hline
     Diag.& $D_{k}$\\
     \hline
       $a1+a2$ & $C_{F}(\frac{1}{\epsilon_{_{_{IR}}}}+\frac{11}{6}+\frac{29}{6}\ln(2))$\\
     \hline
     $a3+a4+a9$
     &$\frac{-3C_{A}}{16\epsilon_{_{_{IR}}}^{2}}+\frac{-32\beta_{0}+15C_{A}-64C_{F}}{32\epsilon_{_{IR}}}-
     \frac{C_{F}(100+172\ln(2)+3\pi^2)}{24}+\frac{C_{A}(75+112\ln(2)+18\pi^2)}{192}
     +\beta_{0}\ln(\frac{\mu^{2}}{4m_{Q}^{2}})$ \\
     \hline
     $a5$  & $\frac{C_{A}}{2}(-\frac{13}{8\epsilon^{2}_{_{IR}}}-\frac{15}{16\epsilon_{{_{IR}}}}-
     \frac{17}{288}+\frac{37}{48}\pi^2+\frac{89}{18}\ln(2))$ \\
     \hline
     $a6$  & 0 \\
      \hline
     $a7$  & $(C_{F}-\frac{C_{A}}{2})(\frac{1}{\epsilon_{IR}}+\frac{1}{2v}\pi^2-
     \frac{5}{3}+\frac{1}{8}\pi^2+\frac{7}{3}\ln(2))$ \\
      \hline
     $a8$  & $-C_{A}(\frac{5}{9}+\frac{10}{9}\ln(2))$ \\
      \hline
\end{tabular}
\end{table}
\end{center}

\subsection{Summation of Real and Virtual Corrections }
Collecting the RC with VC, we obtain the full QCD results at
$\mathcal{O}(\alpha_{s}^{3})$:
\begin{subequations}\label{FullQCDRe}
\begin{align}
(2\textrm{Im}\mathcal{\bar{A}}(^3S_{1}^{[1]}))\Big{|}_{\textrm{pert
QCD}}=\frac{40\alpha_{s}^{3}(\pi^2-9)}{81m_{Q}^{2}},
\end{align}
\begin{eqnarray}
&(2\textrm{Im}\mathcal{\bar{A}}(^3S_{1}^{[8]}))\Big{|}_{\textrm{pert
QCD}}=\frac{\alpha_{s}^{2}}{108m_{Q}^{2}}(5\alpha_{s}(-657+67\pi^2)+N_{f}(36\pi&\nonumber\\
&+\alpha_{s}(36(2C_{F}-C_{A})\frac{\pi^2}{4v}+642-20N_{f}-27\pi^2+72\ln(2)+36\beta_{0}\ln\frac{\mu^{2}}{4m_{Q}^{2}}))),&
\end{eqnarray}
\begin{align}
\begin{array}{c}
(2\textrm{Im}\mathcal{\bar{A}}(^3P_{0}^{[8]}))\Big{|}_{\textrm{pert
QCD}}=\frac{5\alpha_{s}^{2}}{432m_{Q}^{4}}(216\pi+\alpha_{s}(54(2C_{F}-C_{A})\frac{\pi^2}{v}+
3032+21\pi^2 \\
+840\ln(2) +216\beta_{0}\ln(\frac{\mu^2}{4m_{Q}^{2}})))
-\frac{4}{9m_{Q}^{2}}B_{F}\alpha_{s}
(3(\textrm{2Im}\mathcal{A}(^3S_{1}^{[8]}))^{\textrm{Born}}\frac{F_{\epsilon}}{\pi\epsilon}
+\frac{29N_f}{3m_Q^2}\alpha_{s}^{2}),
\end{array}
\end{align}
\begin{align}
\begin{array}{c}
(2\textrm{Im}\mathcal{\bar{A}}(^3P_{1}^{[8]}))\Big{|}_{\textrm{pert
QCD}}=\frac{5\alpha_{s}^{2}(1369-138\pi^2)}{216m_{Q}^{4}}
\\-\frac{4}{9m_{Q}^{2}}B_{F}\alpha_{s}
(3(\textrm{2Im}\mathcal{A}(^3S_{1}^{[8]}))^{\textrm{Born}}\frac{F_{\epsilon}}{\pi\epsilon}
+\frac{8N_f}{3m_Q^2}\alpha_{s}^{2}),
\end{array}
\end{align}
\begin{align}
\begin{array}{c}
(2\textrm{Im}\mathcal{\bar{A}}(^3P_{2}^{[8]}))\Big{|}_{\textrm{pert
QCD}}=\frac{\alpha_{s}^{2}}{216m_{Q}^{4}}(144\pi+\alpha_{s}(36(2C_{F}-C_{A})\frac{\pi^2}{v}+4187-258\pi^2\\
+336\ln(2)+144\beta_{0}\ln(\frac{\mu^2}{4m_{Q}^{2}})))
-\frac{4}{9m_{Q}^{2}}B_{F}\alpha_{s}
(3(\textrm{2Im}\mathcal{A}(^3S_{1}^{[8]}))^{\textrm{Born}}\frac{F_{\epsilon}}{\pi\epsilon}
+\frac{58N_f}{15m_Q^2}\alpha_{s}^{2}),
\end{array}
\end{align}
\begin{align}
(2\textrm{Im}\mathcal{\bar{A}}(^3D_{1}^{[1]}))\Big{|}_{\textrm{pert
QCD}}=
\frac{32}{3m_{Q}^{6}}B_{F}\Phi_{2}F_{\epsilon}\pi\alpha_{s}^{3}\mu^{4\epsilon}
(-\frac{608}{135\epsilon}+\frac{-7744+1605\pi^2}{16200})
\end{align}
\begin{align}
(2\textrm{Im}\mathcal{\bar{A}}(^3D_{2}^{[1]}))\Big{|}_{\textrm{pert
QCD}}=
\frac{32}{3m_{Q}^{6}}B_{F}\Phi_{2}F_{\epsilon}\pi\alpha_{s}^{3}\mu^{4\epsilon}
(-\frac{8}{15\epsilon}+\frac{-23024+2125\pi^2}{1800})
\end{align}
\begin{align}
(2\textrm{Im}\mathcal{\bar{A}}(^3D_{3}^{[1]}))\Big{|}_{\textrm{pert
QCD}}=\frac{32}{3m_{Q}^{6}}B_{F}\Phi_{2}F_{\epsilon}\pi\alpha_{s}^{3}\mu^{4\epsilon}
(-\frac{32}{15\epsilon}+\frac{-28656+2645\pi^{2}}{6300})
\end{align}
\end{subequations}
There are still infrared divergences and Coulomb singularities in
some of the expressions above. As explained in Ref.\cite{BBL2}, the
infrared divergence comes from the soft gluon emission of heavy
quarks, and the Coulomb singularity reflects the behavior of heavy
quarks in the potential region. In next section, both of them will
be repeated precisely when doing the NLO corrections for NRQCD
matrix elements in the corresponding regions.

As mentioned above, the $S$- and  $P$-wave subprocesses have been
studied by Petrelli $et al.$\cite{maltoni}. In their paper, the soft
and collinear singularities are separated with the help of eikonal
approximation and Altarelli-Pasrisi splitting functions, then they
calculate the finite part in 4 dimension. In this paper, we
recalculate them in $D$ dimension directly as a cross check, and get
the same results. The $D$-wave subprocesses have also  been considered
in Refs.\cite{L.Berg,G.Belanger} but they did the calculations in
4 dimension, and regularized the infrared divergence with the
binding energy.
%And our expressions agree with theirs in 4-Dimension.

\section{NRQCD Result and Operator Evolution Equations}

There are three typical energy scales in the heavy quarkonium system,
related to the small parameter $v$. They are $m_{Q}$ (the heavy-quark mass),
$m_{Q}v$ (the typical momentum of heavy quarks in heavy
quarkonium), and $m_{Q}v^{2}$ (the binding energy).
%$m_{Q}$ is a kinematic scale and the other two are dynamical quantities.
Then, there are three dynamical regimes in the NRQCD effective
theory, in which either the heavy-quark or the gluon is on mass
shell, and they are
\begin{equation}
\begin{array}{l}
{\textrm{soft regime}: \qquad\; A^{\mu}_{s}:\quad k_{0}\sim
|\vec{k}|\sim m_{Q}v,\qquad \quad\Psi_{s}:T\sim
|\vec{p}|\sim m_{Q}v}\\
{\textrm{potential regime}: A^{\mu}_{p}:\quad k_{0}\sim m_{Q}v^{2},
|\vec{k}|\sim m_{Q}v,\; \Psi_{p}:T\sim m_{Q}v^{2},\;
|\vec{p}|\sim m_{Q}v}\\
{\textrm{ultrasoft regime}\,: A^{\mu}_{u}:\quad k_{0}\sim
|\vec{k}|\sim m_{Q}v^{2},}
\end{array}
\end{equation}
where $k_{\nu}$ and $p_{\nu}$ are the momenta of the gluon field and
heavy-quark field, respectively, and
$T=p_{0}-m_{Q}=\frac{\vec{p}^{2}}{2m_{Q}}+\mathcal{O}(v^{4})$.
Because there are more than one regimes in the nonrelativistic
system, matching the production and annihilation of external
heavy-quark and antiquark pairs at certain order in $v$ can not be
manifest, though the power counting rule, velocity scaling rule, of
operators in NRQCD is simple. This problem has been addressed in
several papers
\cite{Labelle,Luke,Grinstein,Luke1,Pineda,Beneke,Pineda1,Griebhammer},
and the matching prescriptions based on dimensional regularization
in NRQCD were also clarified. Furthermore, the potential NRQCD
(pNRQCD) effective theory was proposed by introducing the potential
to manage the nonperturbative effect in
Ref.\cite{Brambilla:1999xf}. The NRQCD Feynman rules for propagators
in the Coulomb gauge\cite{Griebhammer}, are shown in Fig.[6], where
$\delta_{tr}^{ij}=\delta_{ij}-\frac{k^{i}k^{j}}{|\mathbf{k}|^{2}}$.
At leading order in $v^2$, the interaction term of the NRQCD Lagrangian
, where the antiheavy terms are neglected, is
$ig_{s}\psi^{\dagger}(A_{0}+\frac{(\mathbf{A}\cdot\boldsymbol{\nabla}+
\boldsymbol{\nabla}\cdot\mathbf{A})}{2m_{Q}})\psi$. And it turns to
be
$ig_{s}\psi^{\dagger}(A_{0}+\frac{\mathbf{A}\cdot\boldsymbol{\nabla}}{m_{Q}})\psi$,
for $\boldsymbol{\nabla}\cdot\mathbf{A}=0$ in the Coulomb gauge. The
Feynman rules for vertex can be read directly and are listed in
Fig.[7]\footnote{The Feynman rules are the same for the
corresponding interaction terms in different regimes  though their
power counting may be not.}. The Feynman rules for anti-heavy quark
can be gotten by charge-conjugation symmetry.
\begin{figure}
\begin{center}
\includegraphics[scale=0.65]{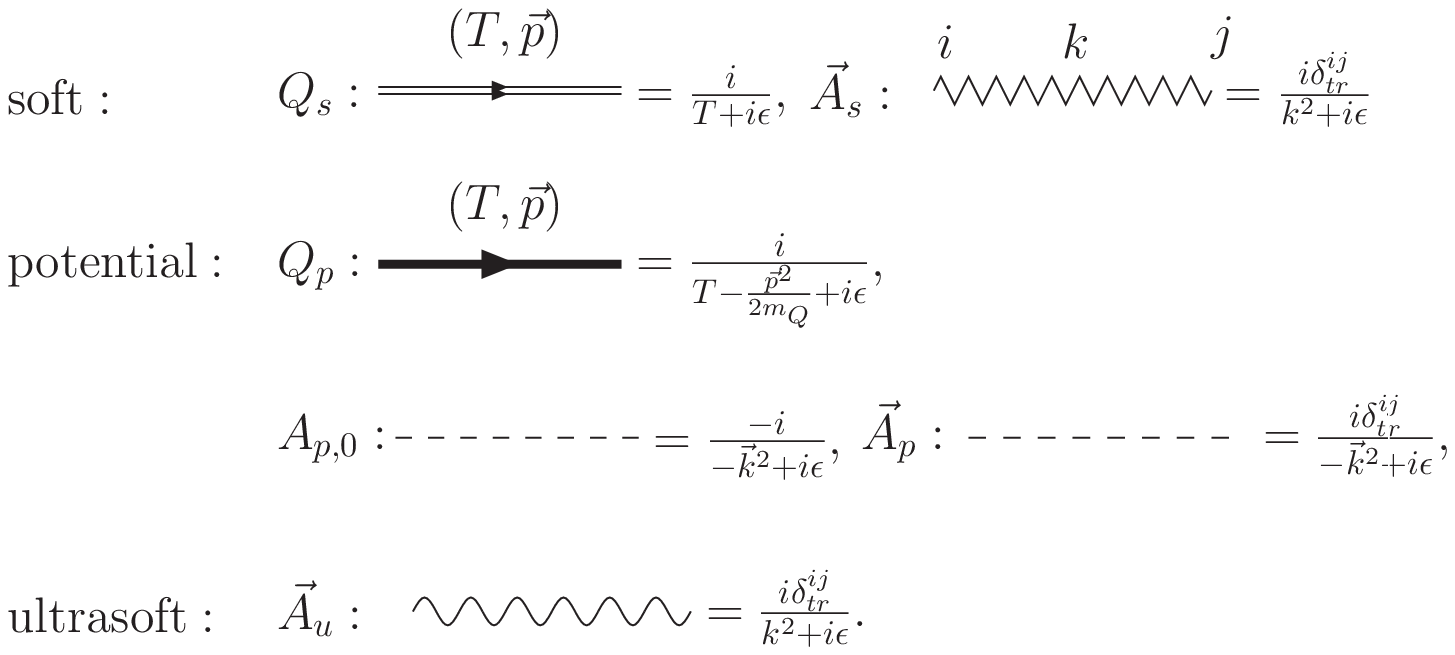}
\caption{NRQCD Feynman rules for heavy-quark and gloun propagators
in different regimes. }
\end{center}
\end{figure}

\begin{figure}
\begin{center}
\includegraphics[scale=0.65]{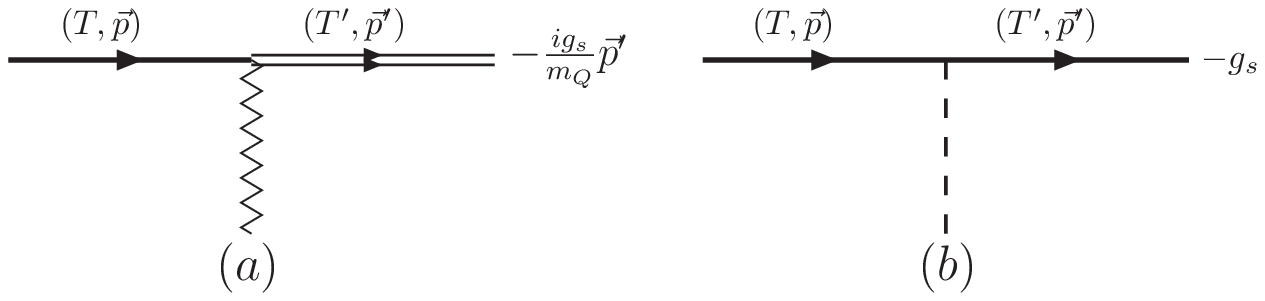}
\caption{NRQCD Feynman rules for heavy-quark gluon vertex. }
\end{center}
\end{figure}

Since the short-distance coefficients are obtained by matching full
QCD results with NRQCD results, we only need to calculate the real
parts  of the matrix elements. Figure 8 gives the LO Feynman diagram.
At NLO in $\alpha_{s}$, when the inner gluon line joints two
incoming or outgoing quark lines, a nonvanishing real part only
appears in the potential region. When the inner gluon line connects
with one incoming quark line and one outgoing quark line, the power
counting rules\cite{Griebhammer} tell us that the soft region will
provide the leading order contribution in $v$. The external self-energy
diagrams are dropped to be in accordance with the renormalization
scheme in full QCD calculation. Then we only need to calculate, two
class, six NLO Feynman diagrams shown in Fig.[9].
\begin{figure}
\begin{center}
\includegraphics[scale=0.5]{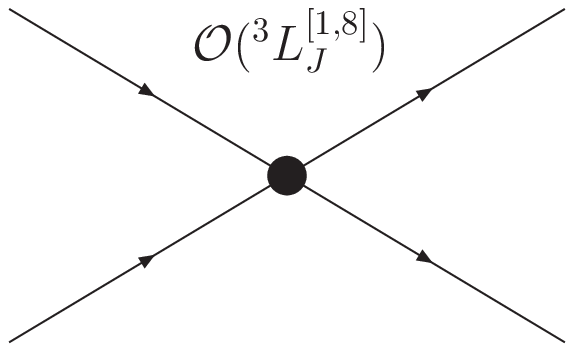}
\caption{NRQCD Feynman Diagram for LO matrix elements}
\end{center}
\end{figure}

\begin{figure}
\begin{center}
\includegraphics[scale=0.5]{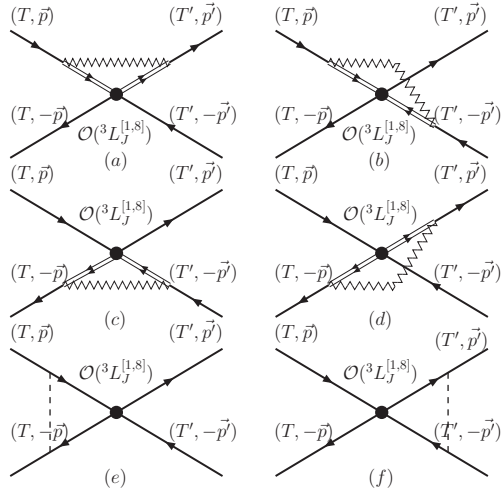}
\caption{NRQCD Feynman Diagrams for NLO matrix elements }
\end{center}
\end{figure}

For convenience, we present the detailed NLO corrections of the
$P$-wave octet matrix elements,
$\langle\mathcal{O}(^3P_{J}^{[8]})\rangle$, which are more
representative than the $S$-wave ones. The LO result
$\langle\mathcal{O}(^3P_{J}^{[8]})\rangle_{\textrm{Born}}$ is
trivial. Using the Feynman rules for propagators in the soft regime and
vertices, the loop integral of diagram (a) reads
\begin{equation}
I_{a}=\frac{ig_{s}^{2}}{m_{Q}^{2}}\int \frac{d^{D}k}{(2\pi)^{D}}
\frac{\mathbf{p}\cdot\mathbf{p'}-(\mathbf{p}\cdot\mathbf{k})(\mathbf{p'}\cdot\mathbf{k})/\mathbf{k}^{2}}
{k_{0}^{2}-\mathbf{k}^{2}+i\epsilon}\frac{1}{k_{0}+i\epsilon}\frac{1}{k_{0}+i\epsilon}
\end{equation}
After performing the contour integrating of
$k_{0}=|\mathbf{k}|-i\epsilon$,
\begin{equation}
I_{a}=\frac{g_{s}^{2}}{2m_{Q}^{2}}\int
\frac{d^{D-1}\mathbf{k}}{(2\pi)^{D-1}}
\frac{\mathbf{p}\cdot\mathbf{p'}-(\mathbf{p}\cdot\mathbf{k})(\mathbf{p'}\cdot\mathbf{k})/\mathbf{k}^{2}}
{|\mathbf{k}|^{3}},
\end{equation}
which is both infrared and ultraviolet divergent. In the dimension
regularization scheme the result is
\begin{equation}
I_{a}=\frac{\alpha_{s}}{3\pi m_{Q}^2}
(\frac{1}{\epsilon_{UV}}-\frac{1}{\epsilon})\mathbf{p}\cdot\mathbf{p'}
\end{equation}
The integrals of (b-d) in Fig.[9] could be calculated in the same
way:
\begin{equation}
I_{b-d}=\frac{\alpha_{s}}{3\pi m_{Q}^2}
(\frac{1}{\epsilon_{UV}}-\frac{1}{\epsilon})\mathbf{p}\cdot\mathbf{p'}
\end{equation}
The loop integral of diagram (e) in potential regime could be
written down similarly:
\begin{equation}
I_{e}=-ig_{s}^{2}\int \frac{d^D
k}{(2\pi)^{D}}\frac{1}{\mathbf{k}^{2}}
\frac{1}{T+k_{0}-\frac{(\mathbf{p}+\mathbf{k})^2}{2m_{Q}}+i\epsilon}\;
\frac{1}{T-k_{0}-\frac{(\mathbf{p}+\mathbf{k})^2}{2m_{Q}}+i\epsilon}
\end{equation}
where $T=\frac{|\mathbf{p}|^{2}}{2m_{Q}}$. When $k_{0}$ is integrated out:
\begin{equation}
I_{e}=g_{s}^{2}m_{Q}\int
\frac{d^{D-1}\mathbf{k}}{(2\pi)^{D-1}}\frac{1}{\mathbf{k}^{2}}
\frac{1}{\mathbf{k}^{2}+2\mathbf{p}\cdot\mathbf{k}-i\epsilon}
\end{equation}
This could be done in $D$ dimension directly. Regularize the Coulomb
singularity by introducing $v=\frac{|\mathbf{p}|}{m_{Q}}$, then at
$v^{-1}$ order we have
\begin{equation}
I_{e}=\frac{\alpha_{s}\pi}{4v}(1-\frac{i}{\pi}(\frac{1}{\epsilon}-\ln(\frac{m_{Q}^{2}v^{2}}{\pi\mu^{2}})-\gamma_E))
\end{equation}
The integral of diagram (f) has the same real part but with a plus sign
before the virtual part.

The color structures for diagrams (a), (c) and (b), (d) and (e), (f)
are obtained by the color decomposition listed in the first, second, and
third line below respectively:
\begin{equation}
\begin{array}{c}
{\sqrt{2}T^{a}T^{b}\otimes
T^{b}\sqrt{2}T^{a}=C_{F}\frac{1}{\sqrt{3}}\otimes
\frac{1}{\sqrt{3}}+\frac{-2}{2N_{c}}
\sqrt{2}T^{c}\otimes \sqrt{2}T^{c},}\\
{\sqrt{2}T^{a}T^{b}\otimes
\sqrt{2}T^{a}T^{b}=C_{F}\frac{1}{\sqrt{3}}\otimes
\frac{1}{\sqrt{3}}+\frac{N_{c}^{2}-2}{2N_{c}} \sqrt{2}T^{c}\otimes
\sqrt{2}T^{c},}\\
{T^{b}\sqrt{2}T^{a}T^{b}\otimes\sqrt{2}T^{a}
=(C_{F}-\frac{1}{2}C_{A})\sqrt{2}T^{c}\otimes\sqrt{2}T^{c}}
\end{array}
\end{equation}

Combining the integrals with the according color factors and summing
them over, we obtain the NLO NRQCD corrections for the $P$-wave octet
operator matrix elements, which are UV divergent and need to be
renormalized:
\begin{equation}
\begin{array}{c}\label{mixing}
{\langle\mathcal{O}^{0}(^3P_{J}^{[8]})\rangle_{NLO}=
\{(1+\frac{\alpha_{s}\pi}{2v}(C_{F}-\frac{1}{2}C_{A}))\sqrt{2}T^{c}\otimes\sqrt{2}T^{c}+}\\
{\frac{4\alpha_{s}}{3\pi
m_{Q}^{2}}(\frac{1}{\epsilon_{_{UV}}}-\frac{1}{\epsilon})
[C_{F}\frac{1}{\sqrt{3}}\otimes\frac{1}{\sqrt{3}}
+B_{F}\sqrt{2}T^{c}\otimes
\sqrt{2}T^{c}]\mathbf{p}\cdot\mathbf{p'}\}
\langle\mathcal{\bar{O}}(^3P_{J})\rangle}_{\textrm{Born}}
\end{array}
\end{equation}
where $\mathcal{O}(^3P_{J}^{[8]})=\mathcal{\bar{O}}(^3P_{J})
\sqrt{2}T^{a}\otimes\sqrt{2}T^{a}$,
$B_{F}=\frac{N_{c}^{2}-4}{4N_{c}}$ and the superscript "0" means the
matrix elements of the bare operators. As expected, at NLO the
$P$-wave octet operators are mixed with the $D$-wave singlet and octet
ones, and with the help of Eq.(\ref{CJPD}), they could be
reexpressed as
\begin{equation}
\begin{array}{c}
{\langle\mathcal{O}^{0}(^3P_{J}^{[8]})\rangle_{NLO}=
(1+\frac{\alpha_{s}\pi}{2v}(C_{F}-\frac{1}{2}C_{A}))\langle\mathcal{O}^{0}(^3P_{J}^{[8]})\rangle_{\textrm{Born}}}
\\
{+\frac{4\alpha_{s}C_{J,J'}}{3\pi
m_{Q}^{2}}(\frac{1}{\epsilon_{_{UV}}}-\frac{1}{\epsilon})(
C_{F}\langle\mathcal{O}^{0}(^3D_{J'}^{[1]})\rangle_{\textrm{Born}}
+B_{F}\langle\mathcal{O}^{0}(^3D_{J'}^{[8]})\rangle_{\textrm{Born}})},
\end{array}
\end{equation}
where $C_{J,J'}$ are defined in Eq.\ref{CJ} and for $J'=1, 2, 3$,
$C_{0,J'}$=$\frac{(D-2)(D+1)}{2(D-1)^{2}}, 0 , 0$;
$C_{1,J'}$=$\frac{(D+1)}{4(D-1)}, \frac{3}{4}, 0$;
$C_{2,J'}$=$\frac{(D-3)^2}{4(D-1)^2}, \frac{1}{4}, 1$. The $P$-wave
operators are mixed with the $D$-wave operators at NLO in
$\alpha_{s}$. But the NLO NRQCD corrections of $D$-wave operators are
related to the relativistic corrections of $P$-wave operators. Then at
leading-order of $v^{2}$, the renormalization transformations of
those operators in $\overline{MS}$ scheme are
\begin{eqnarray}\label{renormalization}
\left(\begin{array}{c} \mathcal{O}^{0}(^3P_{J}^{[8]})\\
\mathcal{O}^{0}(^3D_{J'}^{[1]})
\end{array}\right)=\left(\begin{array}{cc}
1&C_{JJ'}^{^{[1]}}(\frac{1}{\epsilon_{_{UV}}}+\ln4\pi-\gamma_{E})\\
0&1
\end{array}\right)
\left(\begin{array}{c} \mathcal{O}^{R}(^3P_{J}^{[8]})\\
\mathcal{O}^{R}(^3D_{J'}^{[1]})
\end{array}\right)
\end{eqnarray}
where $C_{JJ'}^{^{[1]}}=C_{J,J'}\frac{4\alpha_{s}C_{F}}{3\pi
m_{Q}^{2}}$. When the operators are mixed with each other, the
renormalization constants $Z^{^,}s$ are not numbers but matrices. The
$D$-wave octet operators are also dropped, for they do not appear in
full QCD calculations. The matrix elements of the renormalized
operators $\mathcal{O}^{R}(^3P_{J}^{[8]})$ at NLO are now UV finite,
but there are still $\frac{1}{\epsilon}$ poles in $D$-wave terms,
Eq.(\ref{ReOperator}), and Coulomb singularities, which will absorb
the infrared and Coulomb divergences in full theory:
\begin{equation}
\begin{array}{c}\label{ReOperator}
{\langle\mathcal{O}^{R}(^3P_{J}^{[8]})\rangle_{NLO}=
(1+\frac{\alpha_{s}\pi}{2v}(C_{F}-\frac{1}{2}C_{A}))\langle\mathcal{O}^{R}(^3P_{J}^{[8]})\rangle_{\textrm{Born}}}
\\
{+\frac{4\alpha_{s}C_{J,J'}(\frac{\mu}{\mu_{\Lambda}})^{2\epsilon}}{3\pi
m_{Q}^{2}}(-\frac{1}{\epsilon}-\ln 4\pi+\gamma_{E})(
C_{F}\langle\mathcal{O}^{R}(^3D_{J'}^{[1]})\rangle_{\textrm{Born}}
+B_{F}\langle\mathcal{O}^{R}(^3D_{J'}^{[8]})\rangle_{\textrm{Born}})}.
\end{array}
\end{equation}
where $\mu_{\Lambda}$ is the renormalization scale. The matrix
elements of the $S$-wave octet operator at NLO could be computed in the same way:

\begin{equation}
\begin{array}{c}
{\langle\mathcal{O}^{R}(^3S_{1}^{[8]})\rangle_{NLO}=
(1+\frac{\alpha_{s}\pi}{2v}(C_{F}-\frac{1}{2}C_{A}))\langle\mathcal{O}^{R}(^3S_{1}^{[8]})\rangle_{\textrm{Born}}}
\\
{+\frac{4\alpha_{s}(\frac{\mu}{\mu_{\Lambda}})^{2\epsilon}}{3\pi
m_{Q}^{2}}(-\frac{1}{\epsilon}-\ln 4\pi+\gamma_{E})
(C_{F}\langle\mathcal{O}^{R}(^3P_{J}^{[1]})\rangle_{\textrm{Born}}
+B_{F}\langle\mathcal{O}^{R}(^3P_{J}^{[8]})\rangle_{\textrm{Born}})}.
\end{array}
\end{equation}

The matrix elements of the $S$-wave singlet operator and $D$-wave singlet
operators at NLO do not need to be calculated , for their LO
short-distance coefficients are already at $\mathcal{O}(\alpha_{s}^{3})$.

Multiply the matrix elements with the short-distance coefficients,
we obtain the NRQCD result at NLO in $\alpha_{s}$:
\begin{subequations}\label{NRQCD}
\begin{equation}
(2Im\mathcal{A}(^{3}
S_{1}^{[1]}))_{NRQCD}=\frac{2Imf(^3S_{1}^{[1]})}{m_{Q}^{2}}
\langle\mathcal{O}(^3S_{1}^{[1]})\rangle^{R}_{\textrm{Born}}
\end{equation}
\begin{equation}
(2Im\mathcal{A}(^{3}
S_{1}^{[8]}))_{NRQCD}=\frac{2Imf(^3S_{1}^{[8]})}{m_{Q}^{2}}
(1+\frac{\alpha_{s}\pi}{2v}(C_{F}-\frac{1}{2}C_{A}))\langle\mathcal{O}(^3S_{1}^{[8]})^{R}_{\textrm{Born}}
\end{equation}
\begin{align}
&(2Im\mathcal{A}(^{3}
P_{0}^{[8]}))_{NRQCD}=\{\frac{2Imf(^3P_{0}^{[8]})}{m_{Q}^{4}}
(1+\frac{\alpha_{s}\pi}{2v}(C_{F}-\frac{1}{2}C_{A}))
&\nonumber\\&-\frac{4\alpha_{s}B_{F}}{3\pi
m_{Q}^{4}}\frac{2Imf(^3S_{1}^{[8]})}{\epsilon}(\frac{4\pi\mu^2}{\mu_{\Lambda}^2})^{\epsilon}\Gamma(1+\epsilon)\}
\langle\mathcal{O}(^3P_{0}^{[8]})\rangle^{R}_{\textrm{Born}}&
\end{align}
\begin{align}
&(2Im\mathcal{A}(^{3}
P_{1}^{[8]}))_{NRQCD}=\{\frac{2Imf(^3P_{1}^{[8]})}{m_{Q}^{4}}
(1+\frac{\alpha_{s}\pi}{2v}(C_{F}-\frac{1}{2}C_{A}))&
\nonumber\\&-\frac{4\alpha_{s}B_{F}}{3\pi
m_{Q}^{4}}\frac{2Imf(^3S_{1}^{[8]})}{\epsilon}(\frac{4\pi\mu^2}{\mu_{\Lambda}^2})^{\epsilon}\Gamma(1+\epsilon)\}
\langle\mathcal{O}(^3P_{1}^{[8]})\rangle^{R}_{\textrm{Born}}&
\end{align}
\begin{align}
&(2Im\mathcal{A}(^{3}
P_{2}^{[8]}))_{NRQCD}=\{\frac{2Imf(^3P_{2}^{[8]})}{m_{Q}^{4}}
(1+\frac{\alpha_{s}\pi}{2v}(C_{F}-\frac{1}{2}C_{A}))&
\nonumber\\&-\frac{4\alpha_{s}B_{F}}{3\pi
m_{Q}^{4}}\frac{2Imf(^3S_{1}^{[8]})}{\epsilon}(\frac{4\pi\mu^2}{\mu_{\Lambda}^2})^{\epsilon}\Gamma(1+\epsilon)\}
\langle\mathcal{O}(^3P_{2}^{[8]})\rangle^{R}_{\textrm{Born}}&
\end{align}
\begin{align}
&(2Im\mathcal{A}(^{3}
D_{1}^{[1]}))_{NRQCD}=\{\frac{2Imf(^3D_{1}^{[1]})}{m_{Q}^{6}}&\nonumber\\
&-\sum_{J}\frac{4\alpha_{s}C_{F}C_{J,1}}{3\pi
m_{Q}^{6}}\frac{2Imf(^3P_{J}^{[8]})}{\epsilon}(\frac{4\pi\mu^2}{\mu_{\Lambda}^2})^{\epsilon}\Gamma(1+\epsilon)\}
\langle\mathcal{O}(^3D_{1}^{[1]})\rangle^{R}_{\textrm{Born}}&
\end{align}
\begin{align}
&(2Im\mathcal{A}(^{3}
D_{2}^{[1]}))_{NRQCD}=\{\frac{2Imf(^3D_{2}^{[1]})}{m_{Q}^{6}}&\nonumber\\
&-\sum_{J}\frac{4\alpha_{s}C_{F}C_{J,2}}{3\pi
m_{Q}^{6}}\frac{2Imf(^3P_{J}^{[8]})}{\epsilon}(\frac{4\pi\mu^2}{\mu_{\Lambda}^2})^{\epsilon}\Gamma(1+\epsilon)\}
\langle\mathcal{O}(^3D_{2}^{[1]})\rangle^{R}_{\textrm{Born}}&
\end{align}
\begin{align}
&(2Im\mathcal{A}(^{3}
D_{3}^{[1]}))_{NRQCD}=(\frac{2Imf(^3D_{3}^{[1]})}{m_{Q}^{6}}&\nonumber\\
&-\sum_{J}\frac{4\alpha_{s}C_{F}C_{J,3}}{3\pi
m_{Q}^{6}}\frac{2Imf(^3P_{J}^{[8]})}{\epsilon}(\frac{4\pi\mu^2}{\mu_{\Lambda}^2})^{\epsilon}\Gamma(1+\epsilon))
\langle\mathcal{O}(^3D_{3}^{[1]})\rangle^{R}_{\textrm{Born}}&
\end{align}
\end{subequations}

Finally, matching the NRQCD results with full QCD results, we get
infrared safe short-distance coefficients at
$\mathcal{O}(\alpha_{s}^{3})$:
\begin{subequations}
\begin{equation}
2\textrm{Im}f(^{3}S_{1}^{[1]})=\frac{40\alpha_{s}^{3}(\pi^{2}-9)}{81},
\end{equation}
\begin{align}
&2\textrm{Im}f(^{3}S_{1}^{[8]})=\frac{\alpha_{s}^{2}}{108}
(36N_{f}\pi+\alpha_{s}(5(-657+67\pi^2)\nonumber\\
&+N_{f}(642-20N_{f}-27\pi^{2}+72ln2)+72\beta_{0}N_{f}ln\frac{\mu}{2m_{Q}})),
\end{align}
\begin{align}
&2\textrm{Im}f(^{3}P_{0}^{[8]})=\frac{5\alpha_{s}^{2}}{1296}
(648\pi+\alpha_{s}(9096-464N_{f}\nonumber\\
&+63\pi^2+2520ln2+1296\beta_{0}ln\frac{\mu}{2m_{Q}}+96N_{f}ln\frac{2m_{Q}}{\mu_{\Lambda}})),
\end{align}
\begin{equation}
2\textrm{Im}f(^{3}P_{1}^{[8]})=\frac{5\alpha_{s}^{3}(4107-64N_{f}-414\pi^2+48N_{f}ln\frac{2m_{Q}}{\mu_{\Lambda}})}{648},
\end{equation}
\begin{align}
&2\textrm{Im}f(^{3}P_{2}^{[8]})=\frac{\alpha_{s}^{2}}{648}
(432\pi+\alpha_{s}(12561-464N_{f}\nonumber\\
&-774\pi^2+1008ln2+864\beta_{0}ln\frac{\mu}{2m_{Q}}+240N_{f}ln\frac{2m_{Q}}{\mu_{\Lambda}})),
\end{align}
\begin{align}
2\textrm{Im}f(^{3}D_{1}^{[1]})=\frac{(321\pi^{2}-8032-29184ln\frac{\mu_{\Lambda}}{2m_{Q}})\alpha_{s}^{3}}{5832},
\end{align}
\begin{align}
2\textrm{Im}f(^{3}D_{2}^{[1]})=\frac{(425\pi^{2}-4816-384ln\frac{\mu_{\Lambda}}{2m_{Q}})\alpha_{s}^{3}}{648},
\end{align}
\begin{align}
2\textrm{Im}f(^{3}D_{3}^{[1]})=\frac{(529\pi^{2}-8688-5376ln\frac{\mu_{\Lambda}}{2m_{Q}})\alpha_{s}^{3}}{2268},
\end{align}
\end{subequations}
The $P$- and $D$-wave short-distance coefficients are
$\mu_{\Lambda}$ dependent, and their $\mu_{\Lambda}$ dependence can
be canceled by the renormalized operator $\mu_{\Lambda}$ dependence.
The $\mu_{\Lambda}$ dependence of the renormalized operators could
be derived out by finding the derivative of both sides of
Eq.(\ref{renormalization}) with respect to $\mu_{\Lambda}$:
\begin{subequations}
\begin{equation}
\frac{d\mathcal{O}^{R}(^3P_{J}^{[8]})}{
d\ln\mu_{\Lambda}}=\sum_{J'}\frac{C_{J,J'}8\alpha_{s}C_{F}}{3\pi
m_{Q}^{2}}\mathcal{O}^{R}(^3D_{J'}^{[1]})
\end{equation}
\begin{equation}
\frac{d\mathcal{O}^{R}(^3S_{1}^{[8]})}{
d\ln\mu_{\Lambda}}=\sum_{J}\frac{8\alpha_{s}B_{F}}{3\pi
m_{Q}^{2}}\mathcal{O}^{R}(^3P_{J}^{[8]})
\end{equation}
\begin{equation}
\frac{d\mathcal{O}^{R}(^3S_{1}^{[1]})}{
d\ln\mu_{\Lambda}}=\sum_{J}\frac{1}{2N_{C}}\frac{8\alpha_{s}}{3\pi
m_{Q}^{2}}\mathcal{O}^{R}(^3P_{J}^{[8]})
\end{equation}
\end{subequations}
Remember, for bare quantities,
$\frac{d\mathcal{O}^{0}(^3L_{J}^{[1,8]})}{d\mu_{\Lambda}}=0$. For a
phenomenological reason, we also give the $\mu_{\Lambda}$ dependence of
$\mathcal{O}(^3S_{1}^{[1]})$, though we do not calculate its NLO
NRQCD corrections. By solving the differential equations, all $S$- and
$P$-wave operators' expectation values in $|H_{J'}\rangle$ states are
related to that of the $D$-wave singlet operators:
\begin{subequations}\label{OEE}
\begin{align}
\langle
H_{J'}|\mathcal{O}^{R}(^3P_{J}^{[8]})(\mu_{\Lambda})|H_{J'}\rangle
=\langle
H_{J'}|\mathcal{O}^{R}(^3P_{J}^{[8]})(\mu_{\Lambda_{0}})|H_{J'}\rangle+\nonumber
\\C_{J,J'}\frac{8C_{F}}{3m_{Q}^2\beta_{0}}
\ln\frac{\alpha_{s}(\mu_{\Lambda_{0}})}{\alpha_{s}(\mu_{\Lambda})}
\langle H_{J'}|\mathcal{O}^{R}(^3D_{J'}^{[1]})|H_{J'}\rangle,
\end{align}
\begin{align}
&{\langle
H_{J'}|\mathcal{O}^{R}(^3S_{1}^{[8]})(\mu_{\Lambda})|H_{J'}\rangle
=\frac{C_{F}B_{F}}{2}(\frac{8}{3m_{Q}^2\beta_{0}}
\ln\frac{\alpha_{s}(\mu_{\Lambda_{0}})}{\alpha_{s}(\mu_{\Lambda})})^{2}
\langle
H_{J'}|\mathcal{O}^{R}(^3D_{J'}^{[1]})|H_{J'}\rangle+}&\nonumber\\
&\sum_{J}{\frac{8B_{F}}{3m_{Q}^2\beta_{0}}
\ln\frac{\alpha_{s}(\mu_{\Lambda_{0}})}{\alpha_{s}(\mu_{\Lambda})}\langle
H_{J'}|\mathcal{O}^{R}(^3P_{J}^{[8]})(\mu_{\Lambda_{0}})|H_{J'}\rangle
+\langle
H_{J'}|\mathcal{O}^{R}(^3S_{1}^{[8]})(\mu_{\Lambda_{0}})|H_{J'}\rangle},&
\end{align}
\begin{align}
&{\langle
H_{J'}|\mathcal{O}^{R}(^3S_{1}^{[1]})(\mu_{\Lambda})|H_{J'}\rangle
=\frac{C_{F}}{4N_{C}}(\frac{8}{3m_{Q}^2\beta_{0}}
\ln\frac{\alpha_{s}(\mu_{\Lambda_{0}})}{\alpha_{s}(\mu_{\Lambda})})^{2}
\langle
H_{J'}|\mathcal{O}^{R}(^3D_{J'}^{[1]})|H_{J'}\rangle+}&\nonumber\\
&\sum_J{\frac{4}{3N_{C}m_{Q}^2\beta_{0}}
\ln\frac{\alpha_{s}(\mu_{\Lambda_{0}})}{\alpha_{s}(\mu_{\Lambda})}\langle
H_{J'}|\mathcal{O}^{R}(^3P_{J}^{[8]})(\mu_{\Lambda_{0}})|H_{J'}\rangle
+\langle
H_{J'}|\mathcal{O}^{R}(^3S_{1}^{[1]})(\mu_{\Lambda_{0}})|H_{J'}\rangle},&
\end{align}
\end{subequations}

In pNRQCD, the $S$-wave color-octet matrix elements for $P$-wave heavy
quarkonium decays are also estimated through operator evolution
equation\cite{Brambilla:2001xy,Brambilla:2002nu}. And the relations
between their results and ours are discussed in our previous
work\cite{Fan:2009cj}, which shows that the two methods are
consistent with each other.

\section{Numerical Result and Discussion}
\subsection{$^{3}D_{J}$ Decay into LH  }
For heavy-quark spin-symmetry, the long-distance matrix elements of
$D$-wave four-fermion operators are equal to each other for different $J$,
and relate to the second derivative of wave functions at the origin:
\begin{equation}
\begin{array}{c}
{\frac{15|R''(0)|^{2}}{8\pi}=\langle H_1|\mathcal{O}(^3D_{1}^{[1]})|H_1\rangle=}\\
{\langle H_2|\mathcal{O}(^3D_{2}^{[1]})|H_2\rangle= \langle
H_3|\mathcal{O}(^3D_{3}^{[1]})|H_3\rangle=H_{D}m_{Q}^{6}}
\end{array}
\end{equation}
The matrix elements of the $P$-wave octet operators and the $S$-wave
singlet as well as octet operators in the corresponding $J'$ states
could be estimated through the resolution of operator evolution equations,
Eq.(\ref{OEE}). When $\mu_{\Lambda_{0}}$ and $\mu_{\Lambda}$ are
separated widely enough, the evaluation terms will be much more important
than the boundary terms labeled with $\mu_{\Lambda_{0}}$. Here we
set $\mu_{\Lambda_{0}}=m_{Q}v$, where $v^2=0.3$ for charmonium and
$v^2=0.1$ for bottomonium, since the NRQCD perturbative calculations
could only hold down to scale of order $m_{Q}v$:

\begin{subequations}
\begin{align}
\langle
H_{J'}|\mathcal{O}^{R}(^3P_{J}^{[8]})(\mu_{\Lambda})|H_{J'}\rangle =
C_{J,J'}\frac{8C_{F}}{3\beta_{0}}
\ln\frac{\alpha_{s}(\mu_{\Lambda_{0}})}{\alpha_{s}(\mu_{\Lambda})}H_{D}m_{Q}^4,
\end{align}
\begin{equation}
\begin{array}{c}
{\langle
H_{J'}|\mathcal{O}^{R}(^3S_{1}^{[8]})(\mu_{\Lambda})|H_{J'}\rangle
=\frac{C_{F}B_{F}}{2}(\frac{8}{3\beta_{0}}
\ln\frac{\alpha_{s}(\mu_{\Lambda_{0}})}{\alpha_{s}(\mu_{\Lambda})})^{2}H_{D}m_{Q}^2},
\end{array}
\end{equation}
\begin{equation}
\begin{array}{c}
{\langle
H_{J'}|\mathcal{O}^{R}(^3S_{1}^{[1]})(\mu_{\Lambda})|H_{J'}\rangle
=\frac{C_{F}}{4N_{C}}(\frac{8}{3\beta_{0}}
\ln\frac{\alpha_{s}(\mu_{\Lambda_{0}})}{\alpha_{s}(\mu_{\Lambda})})^{2}H_{D}m_{Q}^2},
\end{array}
\end{equation}
\end{subequations}
We also assume $\mu_{\Lambda}=\mu$, for the factorization scale
$\mu_{\Lambda}$ in NRQCD also acts as the renormalization scale in
operator renormalization. In the end, we come to the overall
expressions for the LH decay widths of $^{3}D_{J}(J=1,2,3)$ states
to NLO in $\alpha_{s}$ at leading order of $v^{2}$:
\begin{subequations}
\begin{align}
\Gamma(^3D_{1}\rightarrow
\mathrm{LH})=(-5.0\alpha_{s}^{3}(0.167+\ln\frac{\mu}{2m_{Q}})+\nonumber\\
\frac{\alpha_{s}^{2}(15.7+\alpha_{s}(88.5-4.33N_{f})+\alpha_{s}(10.0\beta_{0}-1.32N_{f})\ln\frac{\mu}{2m_{Q}})
\ln\frac{\bar{\alpha}_{s}}{\alpha_{s}}}{\beta_{0}}+\nonumber\\
\frac{1.32\alpha_{s}^{2}(1.57N_{f}-0.278\alpha_{s}(N_{f}-21.4)(N_{f}+0.093)+1.0N_{f}\alpha_{s}\beta_{0}\ln\frac{\mu}{2m_{Q}})
\ln^{2}\frac{\bar{\alpha}_{s}}{\alpha_{s}}}{\beta_{0}^{2}})H_{D}.
\end{align}
\begin{align}
\Gamma(^3D_{2}\rightarrow
\mathrm{LH})=(-0.59\alpha_{s}^{3}(1.62+\ln\frac{\mu}{2m_{Q}})+\nonumber\\
\frac{\alpha_{s}^{2}(1.87+\alpha_{s}(8.14-1.95N_{f})+\alpha_{s}(1.19\beta_{0}-1.32N_{f})\ln\frac{\mu}{2m_{Q}})
\ln\frac{\bar{\alpha}_{s}}{\alpha_{s}}}{\beta_{0}}+\nonumber\\
\frac{1.32\alpha_{s}^{2}(1.57N_{f}-0.278\alpha_{s}(N_{f}-21.4)(N_{f}+0.093)+1.0N_{f}\alpha_{s}\beta_{0}\ln\frac{\mu}{2m_{Q}})
\ln^{2}\frac{\bar{\alpha}_{s}}{\alpha_{s}}}{\beta_{0}^{2}})H_{D}.
\end{align}
\begin{align}
\Gamma(^3D_{3}\rightarrow \mathrm{LH})=(-2.37\alpha_{s}^{3}(0.645+\ln\frac{\mu}{2m_{Q}})+\nonumber\\
\frac{\alpha_{s}^{2}(7.45+\alpha_{s}(30.8-2.55N_{f})+\alpha_{s}(4.74\beta_{0}-1.32N_{f})\ln\frac{\mu}{2m_{Q}})
\ln\frac{\bar{\alpha}_{s}}{\alpha_{s}}}{\beta_{0}}+\nonumber\\
\frac{1.32\alpha_{s}^{2}(1.57N_{f}-0.278\alpha_{s}(N_{f}-21.4)(N_{f}+0.093)+1.0N_{f}\alpha_{s}\beta_{0}\ln\frac{\mu}{2m_{Q}})
\ln^{2}\frac{\bar{\alpha}_{s}}{\alpha_{s}}}{\beta_{0}^{2}})H_{D}
\end{align}
\end{subequations}
where $\bar{\alpha_{s}}=\alpha_{s}(\mu_{\Lambda_{0}})$.

\subsubsection{\textrm{$D$-wave Charmonium} $\psi(1^{3}D_{J})$ LH Decay }
Making a choice of $m_{c}=1.5\textrm{GeV},
\Lambda_{QCD}=390\textrm{MeV},$ $H_{D1}=\frac{15|R''_{D}|^2}{8\pi
m_{c}^6}=0.786\times10^{-3}\textrm{GeV}$\cite{Eichten:1995ch} and $N_{f}=3$
for charmonia, we obtain at $\mu=2m_{c}$:
\begin{equation}\label{numerC}
\Gamma(\psi(1^{3}D_{J})\rightarrow \mathrm{LH})=(435,50,172)
\textrm{keV}\quad \textrm{for}~J=(1,2,3).
\end{equation}
When $\mu=m_{c}$ and the other parameters are fixed, the results
turn to be:
\begin{equation}
\Gamma(\psi(1^{3}D_{J})\rightarrow \mathrm{LH})=(683,42,223)
\textrm{keV}\quad \textrm{for}~J=(1,2,3).
\end{equation}
And the $\mu$ dependence of the decay widths at
$\mathcal{O}(\alpha_{s}^{3})$ is shown in Fig.[10]

\begin{figure}
\begin{center}
\includegraphics[scale=1.0]{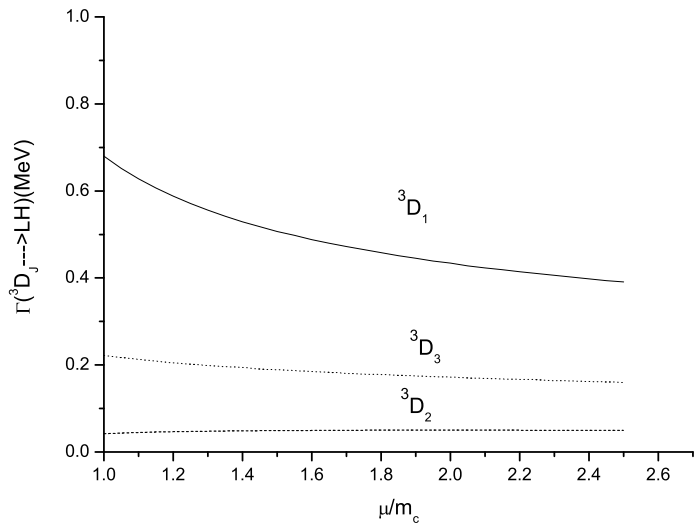}
\caption{Renormalization scale dependence of the decay widths of
charmonium states $1^{3}D_{J}$ to LH  at $\alpha_{s}^{3}$ order.}
\end{center}
\end{figure}

In the potential model, $\psi(1^3D_J)$ can only decay to $3g$ at
$\alpha_{s}^{3}$ order. The infrared divergences are regularized by
$\epsilon_{be}$ the binding energy of the bound states. Accurate to
$\epsilon_{be}$ order, potential model results\cite{L.Berg} are:
\begin{equation}
\Gamma_{C}(^{3}D_{J}\rightarrow \mathrm{LH})=(160,12,68) \textrm{keV},\quad
\textrm{for} ~J=(1,2,3).
\end{equation}
If we reset their parameters the same as ours with
$\alpha_{s}=\alpha_{s}(2m_{c})$, $M=2m_{c}=3.0$GeV,
$|R''_{D}|^2=0.015\textrm{GeV}^{7}$, the potential model predictions
become:
\begin{equation}
\Gamma_{C}(^{3}D_{J}\rightarrow \mathrm{LH})=(240,18,102) \textrm{keV},\quad
\textrm{for} ~J=(1,2,3).
\end{equation}
It could be found that in the $c\bar{c}$ system the NRQCD predictions
are about $2\sim3$ times larger than potential model results. In
leading logarithm approximations\cite{G.Belanger}, the ratios of the
LH decay widths for $J=1,2,3$ are
$\Gamma(^{3}D_{1}):\Gamma(^{3}D_{2}):\Gamma(^{3}D_{3})=\frac{76}{9}:1:4$.
Including the non-negligible corrections to the leading logarithmic
terms\cite{L.Berg}, the ratios turn to be: $40:3:17$. And the
relative ratios predicted by NRQCD at $\mu=2m_c=3.0$GeV and
$\mu=m_c=1.5$GeV are $43:5:17$ and $34:2:11$, respectively.

Much works has been done to predict the mass spectrum of
$\psi(1^3D_J)$; some of the numerical results are collected in
Refs.\cite{Barnes:2003vb,Swanson:2006st}, and some theoretical work
reviews may be found in Ref.\cite{Eichten:2007qx} and references
therein. All the predictions indicated that the masses of
$\psi(1^3D_J)$ are all larger than the threshold of $D\bar{D}$
(about $3730$MeV) , and the center of gravity of $1D$ states
calculated in the Cornell potential\cite{Eichten:1978tg} is
$3815$MeV\cite{Eichten:2002qv}. For its decay to open charm to be
kinematically allowed, $\psi(1^3D_1)$ should not be a narrow state.
It is believed that $\psi(3770)$ is primarily a $1^3D_1$ state with
a small admixture of the $2^3S_1$ state\cite{Ding:1991vu,Rosner:2001nm},
and the latest experimental average of its width is
$\Gamma(\psi(3770))=27.3\pm1.0$MeV\cite{Amsler:2008zz}. But there is
a long-standing puzzle in its non-$D\bar{D}$ decay that the
$\psi(3770)$ decay is not saturated by the $D\bar{D}$
decay\cite{BES1}. A detailed discussion about this problem could be
found in our previous paper \cite{He:2008xb}, and in this paper we
will briefly review it in Sec.V\textbf{B}.

The remaining $J=2$ and $J=3$ states are both expected to be narrow
with different reasons. $\psi(1^3D_2)$ is presumed to lie
between the $D\bar{D}$ and $D\bar{D}^{\ast}$
thresholds\cite{Eichten:2004uh} and is forbidden by parity to decay
into two pseudoscalar $D$ mesons. While narrowness of
$\psi(1^3D_{3})$ in contrast is due to suppression by the $D\bar{D}$
$F$-wave angular momentum
barrier\cite{Barnes:2003vb,Eichten:2004uh}. The principal decay
modes of $\psi(1^3D_2)$ are radiative transition ($\psi(1^3D_2)\to
\gamma\chi_{c1},\gamma\chi_{c2}$), hadronic transition
($\psi(1^3D_2)\to J/\psi\pi\pi$), and LH decay. To $\psi(1^3D_3)$,
these decay modes are also considerable since
$\Gamma(\psi(1^3D_3)\to D\bar{D})$ is predicted to be only about
$0.8$MeV\cite{Eichten:2005ga}, when its mass is $3868$MeV. And the
decay widths predicted by the $C^{3}$
model\cite{Eichten:2004uh,Eichten:2005ga} including the influence of
open-charm channels are
$\Gamma(\psi(1^3D_2)\to\gamma\chi_{c1})=212$keV,
$\Gamma(\psi(1^3D_J)\to\gamma\chi_{c2})=(45,286)$keV, for $J=(2,3)$
and $\Gamma(\psi(1^3D_3)\to D\bar{D})=0.82$MeV at
$m_{\psi(1^3D_2)}=3831$MeV and $m_{\psi(1^3D_3)}=3868$MeV. They also
estimated $\Gamma(\psi(1^3D_J)\to J/\psi\pi\pi)=68\pm15$keV. Using
the numerical values in Eq.(\ref{numerC}), we then roughly predict
that the branching ratios for the LH decay of $\psi(1^3D_J)$ are:
\begin{equation}
 \textrm{Br}(\psi(1^3D_J)\to \mathrm{LH})=13.3\%,13.3\%,\quad\textrm{for}~J=(2,3)
\end{equation}

\subsubsection{$D$-wave Bottomonium $\Upsilon(n^{3}D_{J})$ LH Decay}
Unlike charmonium, $\Upsilon(n^3D_J)$ (for n=1,2) are predicted to
lie below the $B\bar{B}$ flavor threshold, and expected to be quite
narrow, where n is the level number. Some predictions of
$\Upsilon(1^3D_J)$ and $\Upsilon(2^3D_J)$ masses are reviewed in
Ref.\cite{Godfrey:2001vc}. Taking $m_{b}=4.6\textrm{GeV},
\Lambda_{QCD}=340\textrm{MeV},~N_{f}=4$,
$H_{D1}=\frac{15|R''_{1D}|^2}{8\pi
m_{b}^6}=0.401\times10^{-4}\textrm{GeV}$ for $1D$ states, and
$H_{D2}=\frac{15|R''_{2D}|^2}{8\pi
m_{b}^6}=0.750\times10^{-4}\textrm{GeV}$ for $2D$
states\cite{Eichten:1995ch}, at $\mu=2m_b$, we find

\begin{subequations}
\begin{equation}\label{bbn}
\Gamma(\Upsilon(1^{3}D_{J})\rightarrow \mathrm{LH})=(6.91,0.75,2.75)
\textrm{keV}\quad \textrm{for} ~J=(1,2,3)
\end{equation}

\begin{equation}
\Gamma(\Upsilon(2^{3}D_{J})\rightarrow \mathrm{LH})=(12.9,1.40,5.14)
\textrm{keV}\quad \textrm{for} ~J=(1,2,3)
\end{equation}
\end{subequations}
When $\mu=m_{b}$ and the other parameters are unchanged, our
predictions turn to be
\begin{subequations}
\begin{equation}
\Gamma(\Upsilon(1^{3}D_{J})\rightarrow \mathrm{LH})=(7.99,0.60,2.85)
\textrm{keV}\quad \textrm{for} ~J=(1,2,3)
\end{equation}
\begin{equation}
\Gamma(\Upsilon(2^{3}D_{J})\rightarrow \mathrm{LH})=(14.9,1.21,5.33)
\textrm{keV}\quad \textrm{for} ~J=(1,2,3)
\end{equation}
\end{subequations}
The $\mu$ dependence curves of $\Upsilon(1^3D_J)$ and
$\Upsilon(2^3D_J)$ LH decay widths are similar, so only the $n=1$
results are shown at $\mathcal{O}(\alpha_{s}^{3})$ in Fig.[11] as
an illustration.

\begin{figure}
\begin{center}
\includegraphics[scale=1]{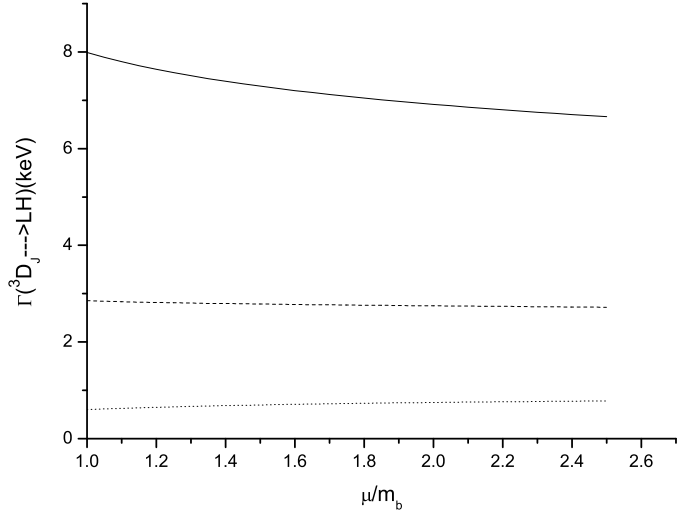}
\caption{Renormalization or factorization scale dependence of
$\Gamma(\Upsilon(1^{3}D_{J}) \to \mathrm{LH} )$ at $\alpha_{s}^{3}$ order.
The solid, dotted and dashed are for $\Upsilon(1^3D_1)$,
$\Upsilon(1^3D_2)$, and $\Upsilon(1^3D_3)$, respectively.}
\end{center}
\end{figure}

In the potential model, B$\acute{\textrm{e}}$langer and
Moxhay\cite{G.Belanger} found, for $J=(1,2,3)$, the leading
logarithmic results are $\Gamma(\Upsilon(1^3D_J)\to
ggg)=(2.2,0.26,1.1)$keV, and a good approximation to the exact phase
space integration given by Bergstr$\ddot{\textrm{o}}$m and
Ernstr$\ddot{\textrm{o}}$m\cite{L.Berg} brings a factor of $2\sim3$
enhancement, and their results are $\Gamma(\Upsilon(1^3D_J)\to
ggg)=(6.3,0.51,2.7)$keV, for $J=(1,2,3)$. If we normalize them using
our inputs at $\mu=2m_b$ and setting $M=2m_b$, potential model
estimations are then
\begin{equation}\label{bbp}
\Gamma(\Upsilon(1^{3}D_{J})\rightarrow \mathrm{LH})=(5.4,0.51,2.3)
\textrm{keV}\quad \textrm{for}~J=(1,2,3),
\end{equation}
which, to some extent, are in agreement with our NRQCD numerical
predictions with $\mu=2m_b$. In the $\Upsilon(1D)$ case, it can be
easily found out that the potential model results are dominated by
the logarithmic terms. And numerically, the NRQCD results are mainly
from the $P-$wave color-octet subprocess contributions. If we
relate the logarithmic term $\ln(1/\epsilon)$ in Eqs.(20-22) of
Ref.\cite{L.Berg} to the evolution term
$\ln\frac{\alpha_s(\mu_{\Lambda_0})}{\alpha_s(\mu_{\Lambda})}$ in
this paper by setting
$\frac{\beta_0\alpha_s}{\pi}\ln(1/\epsilon)=\ln\frac{\alpha_s(\mu_{\Lambda_0})}{\alpha_s(\mu_{\Lambda})}$,
we find the logarithmic terms as well as the $\pi^2$ terms in the
potential model results can be exactly reproduced within the NRQCD
approach. This then provides an alternative way to relate the value
of $\ln\frac{\alpha_s(\mu_{\Lambda_0})}{\alpha_s(\mu_{\Lambda})}$ to
the potential model estimation. Using the inputs $\langle
r\rangle=2.5\mathrm{GeV}^{-1}$ given in Ref.\cite{L.Berg},
$m_b=4.6$GeV, $\alpha_s=0.18$, and $N_f=4$, we get
$\ln\frac{\alpha_s(\mu_{\Lambda_0})}{\alpha_s(\mu_{\Lambda})}=0.58$,
which is consistent with the value we obtained by choosing
$\mu_{\Lambda}=2m_b$ and $\mu_{\Lambda_0}=m_bv_b$.

In Ref.\cite{Kwong:1988ae}, the branching
ratios of some decay modes of $\Upsilon(1^3D_J)$ are summarized in
Table IX, where $\Gamma(\Upsilon(1^3D_1)\to e^{+}e^{-})$ was
calculated in Ref.\cite{Moxhay:1983vu} and
$\Gamma(\Upsilon(1^3D_J)\to\pi\pi)$ was obtained by
Moxhay\cite{Moxhay:1987ch}. Since the LH decay widths of
$\Upsilon(1^3D_J)$ are now calculated in the framework of NRQCD, we
update the theoretical predictions for these ratios in Table IV,
where the numerical results in Eq.(\ref{bbn}) are taken as
estimations for LH decay widths of $\Upsilon(1^3D_J)$. In 2004, the
CLEO Collaboration observed $\Upsilon(1D)$ in the four-photon
cascade process $\Upsilon\to\gamma\chi_{b}(2P)$,
$\chi_{b}(2P)\to\gamma\Upsilon(1D)$,
$\Upsilon(1D)\to\gamma\chi_{b}(1P)$,
$\chi_{b}(1P)\to\gamma\Upsilon(1S)$, followed by $\Upsilon(1S)\to
l^{+}l^{-}$, and the branching ratio is
$\mathcal{B}(\gamma\gamma\gamma\gamma
l^+l^-)_{\Upsilon(1D)}=2.5\pm0.5\pm0.5\cdot10^{-5}$\cite{Bonvicini:2004yj}.
The signals are interpreted as predominantly coming from the
production of $\Upsilon(1^3D_2)$. Small contributions of
$\Upsilon(1^3D_1)$ and $\Upsilon(1^3D_3)$ can not be ruled out. In
the near future, with more accumulated data, all the spin-triplet
$\Upsilon(1^3D_J)$ states may be identified. Unfortunately, the
$D$-wave bottomonium LH decays could not provide a good probe to find
out whether NRQCD is prior to the potential model to describe
the bottomonium system, for the difference between the two theoretical
predictions  is small, unless a very precise measurement is made.

For the n=2 states, no experimental evidence has been observed until
now. To make a theoretical comparison for
$\Gamma(\Upsilon(2^3D_J)\to \mathrm{LH})$, the numerical potential model
predictions are needed.

\begin{center}
\begin{table}
\caption{Summary of the partial widths and branching ratios($B$) for
spin-triplet $b\bar{b}$ $D$-wave states, where
$\Gamma(\Upsilon(1^3D_J)\to \mathrm{LH})$ are predicted by us in
Eq.(\ref{bbn}), and the decay widths of the other modes are the same
as those in Table IX of Ref.\cite{Kwong:1988ae} }
\begin{tabular}{|c|c|c|c|}
     \hline
      Level & Final state & Width (keV)&\quad $B~(\%)\quad$\\
     \hline
       \qquad$\Upsilon(1^3D_1)\qquad$ &\qquad $\gamma+\chi_b(1^3P_0)$\qquad\qquad & 21.4  &53.1\\
     \hline
                          & $\gamma+\chi_b(1^3P_1)$ & 11.3  &28.1\\
     \hline
                          & $\gamma+\chi_b(1^3P_2)$ & 0.58  &1.44\\
     \hline
                          & $\mathrm{LH}$  & 6.91  &17.2\\
     \hline
 & $\Upsilon\pi\pi$ & 0.07  &0.17\\
     \hline
 & $e^+e^-$ & 0.0015  &0.0037\\
     \hline
 &  all &  40.3  &   100   \\
     \hline
       $\Upsilon(1^3D_2)$ &$\gamma+\chi_b(1^3P_1)$ & 22.0  &77.1\\
     \hline

                          & $\gamma+\chi_b(1^3P_2)$ &5.7  &20.0\\
     \hline
                          & $\mathrm{LH}$  & 0.75 &2.63\\
     \hline
 & $\Upsilon\pi\pi$ & 0.07  &0.25\\
     \hline
&  all &  28.5 &   100   \\
     \hline
       $\Upsilon(1^3D_3)$ &$\gamma+\chi_b(1^3P_2)$ & 24.3  &89.6\\
     \hline
                          & $\mathrm{LH}$  & 2.75 &10.1\\
     \hline
 & $\Upsilon\pi\pi$ & 0.07  &0.26\\
     \hline
&  all &  27.1 &   100   \\
     \hline
\end{tabular}
\end{table}
\end{center}

\subsection{LH Decay of $\psi(3770)$}

Recently, BES reported\cite{BES2,BES3,BES4} that the branching ratio of
the $\textrm{non}-D \bar{D}$ decay of $\psi(3770)$ is about $15\%$.
While the corresponding data of CLEO\cite{CLEO} imply zero. The
total width $\Gamma(\psi(3770))$ is $23.0\pm2.7$~MeV
\cite{Yao:2006px}\footnote{In this subsection, we still cite PDG06
data, to be in consistent with our analysis in Ref.\cite{He:2008xb}
}, and the hadronic and E1 radiative transitions contribute about
only 350-400~keV and 1.5-1.8\% to the decay width and the branching
ratio of non-$D\bar D$ decay mode, respectively. To clarify this
puzzle, the annihilation decay of $\psi(3770)$, i.e. $\psi(3700)\to
\mathrm{LH}$, is considered in our previous paper\cite{He:2008xb}, where
$\psi(3770)$ is taken as a $D$-wave dominated state with a small
admixture of the $2S$ state. We found when the annihilation decay is
included, $\Gamma(\psi(3770)\to \mathrm{non}-D\bar{D})$ is  $1.15\sim
~1.20\textrm{MeV}$, corresponding to branching ratio of about $5\%$.

 In the above sections, the short-distance coefficients
and long-distance matrix elements of $\psi(1^3D_1)$ have been given
in detail. Now, we show how to get the $S-D$ mixing term. The typical
Feynman diagram for interference between the color-singlet $^3S_1$ and
$^3D_1$ is shown in Fig.[12]. The interference between other Fock
states of $S$-wave and $D$-wave are suppressed by $\alpha_{s}$ or
$v^{2}$. For example, the interference between two $P$-wave octet
states is of relative $v^{2}$ order. And the $S$-wave singlet
interference is of relative $\alpha_{s}^{2}$ order, since there are
at least two additional gluons in the $S$-wave Fock state of
$^3D_{1}^{[1]}$. In full QCD, the square of the $D$-wave amplitude is
logarithm divergent in phase space integration, and that of $S$-wave
amplitude is finite, therefore, the combination of them will be
finite. Then the short-distance part in Eq.(\ref{S-D}) could be
calculated in 4 dimension:
\begin{equation}\label{S-D}
2\textrm{Im}\mathcal{\bar{A}}((Q\overline{Q})_{^{3}D_{1}}^{[1]}
\rightarrow(Q\overline{Q})_{^{3}S_{1}}^{[1]})=\frac{1}{3} \int
Re[\sum|\mathcal{\overline{M}}((Q\overline{Q})_{^{3}D_{1}}^{[1]}\rightarrow
\mathrm{LH})\mathcal{\overline{M}}^{\ast}((Q\overline{Q})_{^{3}S_{1}}^{[1]}\rightarrow
\mathrm{LH})]d\Phi.
\end{equation}
Taking into account the corresponding long-distance part, we then
obtain the final expression for the mixing term in
Ref.\cite{He:2008xb}:
\begin{equation}
\langle 1^3D_{1}|\mathrm{LH}\rangle\langle \mathrm{LH}|2^3S_{1}\rangle=
\frac{5\alpha_{s}^{3}(-240+71\pi^2)}{324m_{c}^4}
\frac{R_{2S}(0)}{\sqrt{4\pi}}\sqrt{\frac{1}{8\pi}}R_{1D}''(0).
\end{equation}

\begin{figure}
\begin{center}
\includegraphics[scale=0.5]{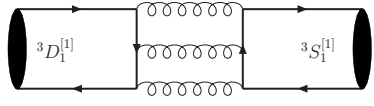}
\caption{QCD Feynman Diagram for the S-D mixing term.}
\end{center}
\end{figure}
\section{Summary}

In this paper, in the framework of NRQCD we study the light hadron
(LH) decays of the spin-triplet ($S$=1) $D$-wave heavy quarkonia. For
completeness, the short-distance coefficients of all Fock states in
the $^3D_J(J=1,2,3)$ quarkonia including $D$-wave color-singlet,
$P$-wave color-octet, and $S$-wave color-singlet and color-octet are
calculated perturbatively at $\alpha_{s}^3$ order. The infrared
divergences in $D$-wave singlet states are absorbed by the $P$-wave
color-octet matrix elements. The operator evolution equations of the
four-fermion operators are also derived and are used to estimate the
numerical values of the long-distance matrix elements. We find that
for the $c\bar{c}$ system, the LH decay widths of $\psi(1^3D_J)$
predicted by NRQCD is about $2\sim3$ times larger than the
phenomenological potential model results, while for the $b\bar{b}$
system the two theoretical estimations of
$\Gamma(\Upsilon(1^3D_J)\to \mathrm{LH})$  are in coincidence with each
other.

\section{Acknowledgement}
This work was supported by the National Natural Science Foundation
of China (No 10675003, No 10721063) and the Ministry of Science and
Technology of China (No 2009CB825200). Zhi-Guo He is currently
supported by the CPAN08-PD14 contract of the CSD2007-00042
Consolider-Ingenio 2010 program, and by the FPA2007-66665-C02-01/
project (Spain).


\begin{thebibliography}{99}

%\cite{Barbieri:1976fp}
\bibitem{Barbieri:1976fp}
  R.~Barbieri, R.~Gatto and E.~Remiddi,
  %``Singular Binding Dependence In The Hadronic Widths Of 1++ And 1+- Heavy
  %Quark Anti-Quark Bound States,''
  Phys.\ Lett.\  B {\bf 61}, 465 (1976);
  %%CITATION = PHLTA,B61,465;%%
 %\cite{Barbieri:1979iy}
%\bibitem{Barbieri:1979iy}
  R.~Barbieri, M.~Caffo and E.~Remiddi,
  %``Gluon Jets From Quarkonia,''
  Nucl.\ Phys.\  B {\bf 162}, 220 (1980).
  %%CITATION = NUPHA,B162,220;%%

\bibitem{BBL6} R.~Barbieri, M.~Caffo, R.~Gatto and E.~Remiddi, Phys. Lett.
\textbf{B95}, 93 (1980), Nucl. Phys.\textbf{B192}, 61 (1981).

\bibitem{L.Berg} L.~Bergstrom and P.~Ernstrom, Phys. Lett. \textbf{B267}, 111 (1991)

\bibitem{G.Belanger} G.~Belanger and P.~Moxhay, Phys. Lett. \textbf{B199}, 575 (1987)


\bibitem{BBL1} G.T.~Bodwin, E.~Braaten, and G.P.~Lepage,
Phys. Rev. \textbf{D46}, R1914.

\bibitem{BBL2} G.T.~Bodwin, E.~Braaten, and G.P.~Lepage,
Phys. Rev. \textbf{D51}, 1125 (1995); \emph{ibid}.\textbf{D55},
5853(E) (1997) [hep-ph/9407339].


%\cite{Huang:1996fa}
\bibitem{Huang1}
  H.~W.~Huang and K.~T.~Chao,
  %``QCD Radiative Correction to the Hadronic Annihilation Rate of $1~{+-}$
  %Heavy Quarkonium,''
  Phys.\ Rev.\  D {\bf 54}, 3065 (1996);
 {\bf 56}, 7472(E)(1997); {\bf 60}, 079901(E) (1999).
 % [arXiv:hep-ph/9601283].
  %%CITATION = PHRVA,D54,3065;%%


\bibitem{Huang3} Han-Wen Huang and Kuang-Ta Chao, Phys. Rev. \textbf{D54}, 6850 (1996);
{\bf 56}, 1821(E) (1997).

\bibitem{Petrelli}  A.~Petrelli, ~ Phys.Lett. \textbf{B380}, 159 (1996)
[hep-ph 9603439]


\bibitem{maltoni}  A.~Petrelli, M.~Cacciari, M.~Greco, F.~Maltoni and M.L.~Mangano, Nucl. Phys.
\textbf{B514},245 (1998)[hep-ph/9707223 v2].

\bibitem{Huang2} Han-Wen Huang and Kuang-Ta Chao, Phys. Rev. \textbf{D55}, 244 (1997)
[hep-ph 9605362 v3].


%\cite{Brambilla:2004wf}
\bibitem{Brambilla:2004wf}
  N.~Brambilla {\it et al.}  [Quarkonium Working Group],
  %``Heavy quarkonium physics,''
  arXiv:hep-ph/0412158.
  %%CITATION = HEP-PH/0412158;%%


%\cite{Brambilla:2008zg}
\bibitem{Brambilla:2008zg}
  N.~Brambilla, E.~Mereghetti and A.~Vairo,
  JHEP 0608:039,2006 [arXiv:hep-ph/0604190];
  %``Hadronic quarkonium decays at order v^7,''
  Phys.\ Rev.\  D {\bf 79}, 074002 (2009)
  [arXiv:0810.2259].
  %%CITATION = PHRVA,D79,074002;%%




\bibitem{Ku} J.H. K\"{u}hn, J. Kaplan, and E.G. Safiani, Nucl. Phys.
\textbf{B157}, 125 (1979); B. Guberina, J.H. K\"{u}hn, R.D. Peccei,
and R. R\"{u}ckl, Nucl. Phys. \textbf{B174}, 317 (1980).

\bibitem{v4} G.T. Bodwin and A. Petrelli, Phys. Rev.
\textbf{D66}, 094011 (2002) [hep-ph/0205210].

\bibitem{FWT1} S.Tani, Proj. Theor. Phys. \textbf{6}, 267 (1951);
L.L.~Foldy and S.A.~Wouthuysen, Phys. Rev. \textbf{78}, 29 (1950).


%\cite{Klasen:2004tz}
\bibitem{Kniehl}
  M.~Klasen, B.~A.~Kniehl, L.~N.~Mihaila and M.~Steinhauser,
  %``$J/\psi$ plus jet associated production in two-photon collisions at
  %next-to-leading order,''
  Nucl.\ Phys.\  B {\bf 713}, 487 (2005);
  [arXiv:hep-ph/0407014].
  %%CITATION = NUPHA,B713,487;%%
%\cite{Klasen:2004az}
%\bibitem{Klasen:2004az}
%  M.~Klasen, B.~A.~Kniehl, L.~N.~Mihaila and M.~Steinhauser,
  %``$J/\psi$ plus prompt-photon associated production in two-photon collisions
  %at next-to-leading order,''
  Phys.\ Rev.\  D {\bf 71}, 014016 (2005)
  [arXiv:hep-ph/0408280].
  %%CITATION = PHRVA,D71,014016;%%


\bibitem{Labelle} P.~ Labelle, MCGILI-96-23 ~[hep-ph/9608491].

\bibitem{Luke} M.~Luke and A.V.~Manohar, Phys.  Rev. \textbf{D55},
4129 (1997).

\bibitem{Grinstein} B.~Grinstein and I.Z.~Rothstein, Phys. Rev.
\textbf {57}, 78 (1998).


\bibitem{Luke1}  M.~Luke and M.J.~Savage,  Phys. Rev. \textbf{D57},
413 (1998).

\bibitem{Pineda}  A.~Pineda and J.~Soto, Nucl.~Phys. Proc.
Suppl.~\textbf{64},~428 (1998) [hep-ph/9707481].

\bibitem{Beneke}  M.~Beneke and V.A.~Smirnov,
Nucl.~Phys. \textbf{B522},~321 (1998); CERN-TH-97-315,
[hep-ph/9711391].

\bibitem{Pineda1} A.~Pineda and J.~soto, Phys. ~Lett. \textbf{B420},
391 (1998); Phys. Rev. D {\bf 59}, 016005 (1988).

\bibitem{Griebhammer} H.W.~Grie{\ss}hammer,  Phys.~Rev. \textbf{D58},~094027
(1998); [hep-ph/9804251].

%\cite{Brambilla:1999xf}
\bibitem{Brambilla:1999xf}
  N.~Brambilla, A.~Pineda, J.~Soto and A.~Vairo,
  %``Potential NRQCD: An effective theory for heavy quarkonium,''
  Nucl.\ Phys.\  B {\bf 566}, 275 (2000)
  [arXiv:hep-ph/9907240];
  %%CITATION = NUPHA,B566,275;%%
  Phys. Rev. D63, 014023 (2000) [arXiv:hep-ph/0002250].

%\cite{Brambilla:2001xy}
\bibitem{Brambilla:2001xy}
  N.~Brambilla, D.~Eiras, A.~Pineda, J.~Soto and A.~Vairo,
  %``New predictions for inclusive heavy-quarkonium P wave decays,''
  Phys.\ Rev.\ Lett.\  {\bf 88}, 012003 (2001)
  [arXiv:hep-ph/0109130].
  %%CITATION = PRLTA,88,012003;%%

%\cite{Brambilla:2002nu}
\bibitem{Brambilla:2002nu}
  N.~Brambilla, D.~Eiras, A.~Pineda, J.~Soto and A.~Vairo,
  %``Inclusive decays of heavy quarkonium to light particles,''
  Phys.\ Rev.\  D {\bf 67}, 034018 (2003)
  [arXiv:hep-ph/0208019].
  %%CITATION = PHRVA,D67,034018;%%


%\cite{Fan:2009cj}
\bibitem{Fan:2009cj}
  Y.~Fan, Z.~G.~He, Y.~Q.~Ma and K.~T.~Chao,
  %``Predictions of Light Hadronic Decays of Heavy Quarkonium 1D_2 States in
  %NRQCD,''
  Phys.\ Rev.\  D {\bf 80}, 014001 (2009)
  [arXiv:0903.4572 [hep-ph]].
  %%CITATION = PHRVA,D80,014001;%%




%\cite{Eichten:1995ch}
\bibitem{Eichten:1995ch}
  E.~J.~Eichten and C.~Quigg,
  %``Quarkonium Wave Functions At The Origin,''
  Phys.\ Rev.\  D {\bf 52}, 1726 (1995)
  [arXiv:hep-ph/9503356].
  %%CITATION = PHRVA,D52,1726;%%

%\cite{Barnes:2003vb}
\bibitem{Barnes:2003vb}
  T.~Barnes and S.~Godfrey,
  %``Charmonium Options for the X(3872),''
  Phys.\ Rev.\  D {\bf 69}, 054008 (2004)
  [arXiv:hep-ph/0311162].
  %%CITATION = PHRVA,D69,054008;%%

%\cite{Swanson:2006st}
\bibitem{Swanson:2006st}
  E.~S.~Swanson,
  %``The new heavy mesons: A status report,''
  Phys.\ Rept.\  {\bf 429}, 243 (2006)
  [arXiv:hep-ph/0601110].
  %%CITATION = PRPLC,429,243;%%

%\cite{Eichten:2007qx}
\bibitem{Eichten:2007qx}
  E.~Eichten, S.~Godfrey, H.~Mahlke and J.~L.~Rosner,
  %``Quarkonia and their transitions,''
  Rev.\ Mod.\ Phys.\  {\bf 80}, 1161 (2008)
  [arXiv:hep-ph/0701208].
  %%CITATION = RMPHA,80,1161;%%

%\cite{Eichten:1978tg}
\bibitem{Eichten:1978tg}
  E.~Eichten, K.~Gottfried, T.~Kinoshita, K.~D.~Lane and T.~M.~Yan,
  %``Charmonium: The Model,''
  Phys.\ Rev.\  D {\bf 17}, 3090 (1978);
   {\bf 21}, 313(E) (1980).
  %%CITATION = PHRVA,D17,3090;%%

%\cite{Eichten:2002qv}
\bibitem{Eichten:2002qv}
  E.~J.~Eichten, K.~Lane and C.~Quigg,
  %``B-Meson Gateways to Missing Charmonium Levels,''
  Phys.\ Rev.\ Lett.\  {\bf 89}, 162002 (2002)
  [arXiv:hep-ph/0206018].
  %%CITATION = PRLTA,89,162002;%%


%\cite{Ding:1991vu}
\bibitem{Ding:1991vu}
  Y.~B.~Ding, D.~H.~Qin and K.~T.~Chao,
  %``Electric Dipole Transitions Of Psi (3770) And S - D Mixing Between Psi
  %(3686) And Psi (3770),''
  Phys.\ Rev.\  D {\bf 44}, 3562 (1991).
  %%CITATION = PHRVA,D44,3562;%%

%\cite{Rosner:2001nm}
\bibitem{Rosner:2001nm}
  J.~L.~Rosner,
  %``Charmless final states and S-D-wave mixing in the psi'',''
  Phys.\ Rev.\  D {\bf 64}, 094002 (2001)
  [arXiv:hep-ph/0105327].
  %%CITATION = PHRVA,D64,094002;%%


%\cite{Amsler:2008zz}
\bibitem{Amsler:2008zz}
  C.~Amsler {\it et al.}  [Particle Data Group],
  %``Review of particle physics,''
  Phys.\ Lett.\  B {\bf 667}, 1 (2008).
  %%CITATION = PHLTA,B667,1;%%

\bibitem{BES1} G.Rong, D.H.Zhang and J.C.Chen, [hep-ex/0506051]


%\cite{He:2008xb}
\bibitem{He:2008xb}
  Z.~G.~He, Y.~Fan and K.~T.~Chao,
  %``QCD prediction for the non-$D\bar D$ annihilation decay of $\psi(3770)$,''
  Phys.\ Rev.\ Lett.\  {\bf 101}, 112001 (2008)
  [arXiv:0802.1849 [hep-ph]].
  %%CITATION = PRLTA,101,112001;%%

%\cite{Eichten:2004uh}
\bibitem{Eichten:2004uh}
  E.~J.~Eichten, K.~Lane and C.~Quigg,
  %``Charmonium levels near threshold and the narrow state $X(3872) \to
  %\pi^{+}\pi^{-}\jpsi$,''
  Phys.\ Rev.\  D {\bf 69}, 094019 (2004)
  [arXiv:hep-ph/0401210].
  %%CITATION = PHRVA,D69,094019;%%


%\cite{Eichten:2005ga}
\bibitem{Eichten:2005ga}
  E.~J.~Eichten, K.~Lane and C.~Quigg,
  %``New states above charm threshold,''
  Phys.\ Rev.\  D {\bf 73}, 014014 (2006)
  [Erratum-ibid.\  D {\bf 73}, 079903 (2006)]
  [arXiv:hep-ph/0511179].
  %%CITATION = PHRVA,D73,014014;%%

%\cite{Godfrey:2001vc}
\bibitem{Godfrey:2001vc}
  S.~Godfrey and J.~L.~Rosner,
  %``Production of the D-wave b anti-b states,''
  Phys.\ Rev.\  D {\bf 64}, 097501 (2001);
   {\bf 66}, 059902(E) (2002).
 % [arXiv:hep-ph/0105273].
  %%CITATION = PHRVA,D64,097501;%%


%\cite{Kwong:1988ae}
\bibitem{Kwong:1988ae}
  W.~Kwong and J.~L.~Rosner,
  %``D WAVE QUARKONIUM LEVELS OF THE UPSILON FAMILY,''
  Phys.\ Rev.\  D {\bf 38}, 279 (1988).
  %%CITATION = PHRVA,D38,279;%%

%\cite{Moxhay:1983vu}
\bibitem{Moxhay:1983vu}
  P.~Moxhay and J.~L.~Rosner,
  %``Relativistic Corrections In Quarkonium,''
  Phys.\ Rev.\  D {\bf 28}, 1132 (1983).
  %%CITATION = PHRVA,D28,1132;%%

%\cite{Moxhay:1987ch}
\bibitem{Moxhay:1987ch}
  P.~Moxhay,
  %``Hadronic Transitions Of D Wave Quarkonium,''
  Phys.\ Rev.\  D {\bf 37}, 2557 (1988).
  %%CITATION = PHRVA,D37,2557;%%


%\cite{Bonvicini:2004yj}
\bibitem{Bonvicini:2004yj}
  G.~Bonvicini {\it et al.}  [CLEO Collaboration],
  %``First observation of a Upsilon(1D) state,''
  Phys.\ Rev.\  D {\bf 70}, 032001 (2004)
  [arXiv:hep-ex/0404021].
  %%CITATION = PHRVA,D70,032001;%%


%\bibitem{BES2} BES Collaboration, M.Ablikim et al., Phys. Lett. \textbf{B641}, 145 (2006)
%\cite{Ablikim:2006aj}
\bibitem{BES2}
  M.~Ablikim {\it et al.}  [BES Collaboration],
  %``Measurements of the cross sections for e+ e- --> hadrons at 3.650-GeV,
  %3.6648-GeV, 3.773-GeV and the branching fraction for psi(3770) --> non  D
  %anti-D,''
  Phys.\ Lett.\  B {\bf 641}, 145 (2006).
%  [arXiv:hep-ex/0605105].
  %%CITATION = PHLTA,B641,145;%%


%\bibitem{BES3} BES Collaboration, M.Ablikim et al., Phys. Rev. Lett. \textbf{97}, 121801 (2006)
%\cite{Ablikim:2006zq}
\bibitem{BES3}
  M.~Ablikim {\it et al.}  [BES Collaboration],
  %``Measurements of the branching fractions for psi(3770) --> D0 anti-D0,  D+
  %D-, D anti-D and the resonance parameters of psi(3770) and psi(2S),''
  Phys.\ Rev.\ Lett.\  {\bf 97}, 121801 (2006).
%  [arXiv:hep-ex/0605107].
  %%CITATION = PRLTA,97,121801;%%


%\bibitem{BES4} BES Collaboration, M.Ablikim et al., Phys. Rev. \textbf{D76}, 122002 (2007)

%\cite{Ablikim:2007zz}
\bibitem{BES4}
  M.~Ablikim {\it et al.},
  %``Direct measurements of the non-D anti-D cross section sigma psi(3770) $\to$
  %non-D anti-D at Ecm=3.773-GeV and the branching fraction for psi(3770) $\to$
  %non-D anti-D,''
  Phys.\ Rev.\  D {\bf 76}, 122002 (2007).
  %%CITATION = PHRVA,D76,122002;%%



%\bibitem{CLEO} CLEO Collaboration, D.Besson et al., Phys. Rev. Lett. \textbf{96}, 092002 (2006)

%\cite{Besson:2005hm}
\bibitem{CLEO}
  D.~Besson {\it et al.}  [CLEO Collaboration],
  %``Measurement of sigma(e+ e- --> psi(3770) --> hadrons) at E(cm) =
  %3773-MeV,''
  Phys.\ Rev.\ Lett.\  {\bf 96}, 092002 (2006).
%  [Erratum-ibid.\  {\bf 104}, 159901 (2010)]
%  [arXiv:hep-ex/0512038].
  %%CITATION = PRLTA,96,092002;%%



%\cite{Yao:2006px}
\bibitem{Yao:2006px}
  W.~M.~Yao {\it et al.}  [Particle Data Group],
  %``Review of particle physics,''
  J.\ Phys.\ G {\bf 33}, 1 (2006).
  %%CITATION = JPHGB,G33,1;%%





\end{thebibliography}
\end{document}